\documentclass[fleqn,usenatbib]{mnras}

\usepackage{newtxtext,newtxmath}

\usepackage[T1]{fontenc}

\DeclareRobustCommand{\VAN}[3]{#2}
\let\VANthebibliography\thebibliography
\def\thebibliography{\DeclareRobustCommand{\VAN}[3]{##3}\VANthebibliography}

\usepackage{graphicx}	
\usepackage{breakurl}

\newcommand{\kms}{~km s$^{-1}$~}

\newcommand{\WR}{WR~140~}
\newcommand{\WRE}{WR~140}
\newcommand{\dotMyr}{~M$_{\odot}$~yr$^{-1}$~}

\newcommand{\Chandra}{{\it Chandra~}}
\newcommand{\ChandraE}{{\it Chandra}}

\newcommand{\rxte}{{\it RXTE~}}
\newcommand{\rxteE}{{\it RXTE}}

\newcommand{\xspec}{{\sc xspec~}}
\newcommand{\xspecE}{{\sc xspec}}

\title[CSWs in \WRE]
{Colliding stellar wind modelling of the X-ray emission from \WR
}

\author[S.A.Zhekov]{Svetozar A. Zhekov\thanks{
E-mail: szhekov@astro.bas.bg}\\
Institute of Astronomy and National Astronomical Observatory
(Bulgarian Academy of Sciences),\\
72 Tsarigradsko Chaussee Blvd., Sofia 1784, Bulgaria\\
}

\date{}

\pubyear{2020}

\begin{document}
\label{firstpage}
\pagerange{\pageref{firstpage}--\pageref{lastpage}}
\maketitle

\begin{abstract}
We modelled the \Chandra and \rxte X-ray spectra of the massive binary
\WR in the framework of the {\it standard} colliding stellar wind
(CSW) picture.
Models with partial electron heating at the shock fronts
are a better representation of the X-ray data than those with complete
temperature equalization.
Emission measure of the X-ray plasma in the CSW region exhibits
a considerable decrease at orbital phases near periastron. This is
equivalent to variable effective mass-loss rates over the binary
orbit.
At orbital phases near periastron, a considerable X-ray absorption
in {\it excess} to that from the stellar winds  in \WR is present.
The {\it standard} CSW model provides line profiles that in general do
{\it not} match well the observed line profiles of the strong line
features in the X-ray spectrum of \WRE.
The variable effective mass-loss rate could be understood {\it
qualitatively} in CSW picture of clumpy stellar winds where clumps are
efficiently dissolved in the CSW region near apastron but not at
periastron.
However, future development of CSW models with
{\it non-spherically-symmetric}
stellar winds might be needed to get a better correspondence between
theory and observations.

\end{abstract}




\section{Introduction}
Massive stars of early spectral types, Wolf-Rayet (WR) and OB,
possess massive and fast supersonic stellar winds 
(V$_{wind} = 1000-5000$\kms; 
$\dot{M}$ $\sim$ 10$^{-7} -  10^{-4}$ \dotMyr).
If a binary consists of two massive stars, the interaction of their
supersonic winds will result in enhanced X-ray emission of such an 
object, originating from plasma heated in colliding-stellar-wind 
(CSW) shocks, as first proposed by \citet{cherep_76} and
\citet{pri_us_76}. Due to the high velocities of stellar winds in
massive stars, X-rays from such binaries provide direct pieces
of information about the shocked plasma.
Thus, the wind collision in a massive binary could simply serve as
an excellent laboratory to test our understanding of physical
processes in extreme conditions of strong shocks: high temperatures, 
rarefied plasmas etc. To do so, we need to confront our CSW theories 
with good experimental data, that is with X-ray observations of very 
good quality that provide spectra with considerable detail (e.g., with
high spectral resolution, orbital coverage). 

Although the first systematic X-ray survey of WRs with
the {\it Einstein} Observatory showed that  WR$+$O
binaries are the brightest X-ray sources amongst them \citep{po_87},
only the modern X-ray observatories 
launched in the last two decades or so 
provided such high-quality data sets that would allow us to carry out 
the corresponding theoretical CSW studies in great detail
(see \citealt{rauw_naze_16} for a review on the
progress of studies of X-ray 
emission from interacting wind massive binaries
of early spectral types).

  It is worth noting that for efficient testing of our theories, we 
need not only X-ray data with good quality, but we need the binary
and stellar parameters of this studied object to be well constrained 
as well. From such a point of view, \WR is probably the `best' object 
amongst the massive WR binaries.
So, the goal of the current study of the colliding stellar wind 
phenomenon is to carry out a direct comparison of the CSW
model results and high-quality X-ray observations of the massive 
Wolf-Rayet binary \WRE:
a long-time pending study.
We note that a direct confrontation of CSW models and observations of
\WR has been carried out but only for four low-resolution 
(undispersed) {\it ASCA} spectra \citep{zhsk_00}.

Our paper is organized as follows.
In Section~\ref{sec:star}, we provide some basic parameters of \WR
adopted in this study.
In Section~\ref{sec:data}, we describe the archival X-ray
observations of \WRE.
In Section~\ref{sec:modelling}, we give details about the CSW
modelling of the X-ray spectra of \WR and the corresponding results.
In Section~\ref{sec:discussion}, we discuss our results, and we 
present our conclusions in Section~\ref{sec:conclusions}.

\section{The Wolf-Rayet binary \WRE}
\label{sec:star}
The WR star \WR (HD 193793) is a massive binary 
(WC7pd$+$O4-5\footnote{Galactic Wolf Rayet Catalogue;
\url{http://pacrowther.staff.shef.ac.uk/WRcat/index.php}}) 
that is fairly considered the prototype of colliding wind binaries.
It shows variable X-ray, radio and infrared emission over its 
$\sim 7.9$-year orbital period \citep{williams_90}. \WR is also one of
the seven WR binaries which originally defined the group of the 
so-called periodic dust makers 
(\citealt{williams_95}, \citealt{williams_08} and references therein).

Its orbital parameters are well constrained as a result from extensive
studies in the optical and near-infrared spectral domains (e.g.,
\citealt{marchenko_03}, \citealt{fahed_11}, \citealt{monnier_11}).
For this study, we use the orbital parameters of \WR
inclination angle ($i$), eccentricity ($e$), semi-major
axis  from \citet{monnier_11} and we note that the adopted
distance by these authors (1670 pc) differs by less than 2\% from
the {\it Gaia} distance of d $= 1641^{+81}_{-74}$~pc
\citep{bailer_jones_18}
and d $= 1640^{+110}_{-90}$~pc \citep{rc_20}
to this object.

For the stellar wind parameters (velocity and mass loss;
V$_{WR}$, $\dot{M}_{WR}$, V$_{O}$, $\dot{M}_{O}$),
we adopt the corresponding values from \citet{williams_90} as the
mass-loss rates were re-scaled to the {\it Gaia} distance and adopting
a modest volume filling factor of 0.25.
Table~\ref{tab:wr140} summarises the \WR parameters adopted in this
study.

Based on the \citet{monnier_11} ephemeris, we derive that the WR star
is in front at orbital phase $= 0.00319$, while the O star is
in front at orbital phase $= 0.95516$.

\begin{table}

\caption{\WR parameters adopted in this work for use in the CSW model
\label{tab:wr140}}
\begin{center}
\begin{tabular}{lcc}
\hline
\multicolumn{1}{c}{Parameter} &
\multicolumn{1}{c}{Value}  &
\multicolumn{1}{c}{Reference}  \\
\hline
T$_0$ (MJD)               &  46154.8   &  (1) \\
Period (dyas)             &  2896.35   &  (1) \\
$\omega$ (deg)            &  46.8      &  (1) \\
Inclination (deg)         &  119.6 (60.4)      &  (1) \\
Eccentricity              &  0.8964    &  (1) \\
Semi-major axis (au)      &  14.73     &  (1) \\
V$_{WR}$ (\kms)              &  2860   &  (2) \\
$\dot{M}_{WR}$ (\dotMyr)  &  $4.04\times10^{-5}$   &  (2) \\
V$_{O} ($\kms)               &  3200   &  (2) \\
$\dot{M}_{O}$ (\dotMyr)   &  $1.275\times10^{-6}$  &  (2) \\
Distance (pc)             &  1641      &  (4) \\
\hline

\end{tabular}
\end{center}

{\it Note}. 
(1) Orbital parameters are from \citet{monnier_11} (see table 2
therein.
(2) \citet{williams_90} as the mass-loss rates were re-scaled to the
{\it Gaia} distance.
(3) The {\it Gaia} distance from \citep{bailer_jones_18} and
\citep{rc_20}.

\end{table}

\section{Observations and data reduction}
\label{sec:data}
In this study we made use of archive X-ray data from observations 
with the \Chandra and \rxte (Rossi X-ray Timing Explorer) X-ray 
observatories. These data sets provide a very dense coverage of the
orbital period of \WR (\rxteE) and high-resolution X-ray spectra with
good photon statistics (\ChandraE).

\subsection{Observations with \Chandra}
\label{sec:data_cha}
We used seven observations with the \Chandra High-Energy Transmission
Gratings (HETG) carried out in the period 2000 December - 2008 August: 
Obs ID 2337, 2338, 5419, 6286, 6287, 8911 and 9909. 
Following the Science Threads for Grating Spectroscopy in the
{\sc ciao} 4.12\footnote{Chandra Interactive Analysis of Observations
(CIAO), \url{https://cxc.harvard.edu/ciao/}.} data analysis software,
we extracted the \WR X-ray spectra from these data sets (we have 
initially re-processed the data adopting the {\sc ciao 
chandra\_repro} script). In this study, we used the first-order
Medium Energy Grating (MEG) and High Energy Grating (HEG) spectra. We
constructed a total spectrum for observations with orbital phases
that are very close and no spectral change is expected. Thus, we 
end up with four data sets Obs 1 (2337), Obs 2 (2338), Obs 3 (5419, 
6286, 6287) and Obs 4 (8911, 9909) that provide high-quality spectra 
with good spectral resolution. The total number of source counts in
the corresponding MEG and HEG spectra are 
50009, 26064 (Obs 1); 5150, 3833 (Obs 2); 54648, 24854 (Obs 3);
71273, 32324 (Obs 4).
The \Chandra calibration database {\sc calbd} v.4.9.0 was used to
construct the response matrices and the ancillary response files.

A conventional X-ray analysis 
(i.e., using discrete-temperature plasma models, deriving line 
profile parameters etc.)
of some of these data was presented in
\citet{po_05}.

\subsection{Observations with \rxte}
\label{sec:data_rxte}
We used 552 archive \rxte spectra of \WRE. The observations were 
carried out in the period 2000 December - 2011 December being part of
eight observational programmes P60004, P70003, P91002, P92003, P93001,
P94001, P95303 and P96300. The typical exposure times were of 500 - 
2200 s that resulted in good to excellent (700 - 50000 source counts)
undispersed X-ray spectra at energies above 2-3 keV. For the spectral
analysis, we used all the source and background spectra, and response 
files as provided in the \rxte archive\footnote{We used \rxte data 
provided by the High Energy Astrophysics Science Archive Research 
Center (HEASARC), \url{https://heasarc.gsfc.nasa.gov/}.}.

Some of these data (P60004) were presented and discussed in
\citet{po_05} and \citet{rus_11}, 
while results from spectral analysis
of the entire \rxte data set are shown in fig. 2 in  \citet{po_12}.

\begin{figure}
\begin{center}
\includegraphics[width=\columnwidth]{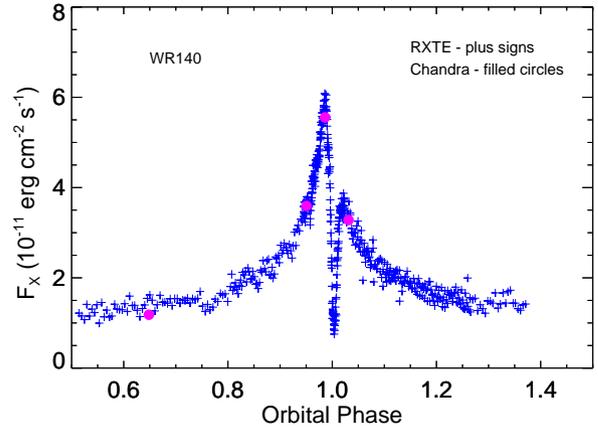}
\end{center}
\caption{The observed X-ray flux in the (3 - 10 keV) energy range from
the \Chandra and \rxte observations of \WRE.
}
\label{fig:rxte_cha_flux}
\end{figure}

Anticipating the results from the CSW modelling of the X-ray spectra
of \WRE, the observed X-ray flux is shown in
Fig.~\ref{fig:rxte_cha_flux}. We see that there is a good
correspondence between the observed flux values from \rxte and
\Chandra observations for similar orbital phases. This gives us
confidence that both spectral data set can be used to test in detail
the CSW picture in this massive binary system.

For the spectral analysis in this study, we made use of 
standard as well as custom models in version 12.10.1 of 
\xspec \citep{Arnaud96}.

\section{Spectral modelling}
\label{sec:modelling}

\subsection{CSW model}
\label{sec:csw}
\begin{figure}
\begin{center}
\includegraphics[width=\columnwidth]{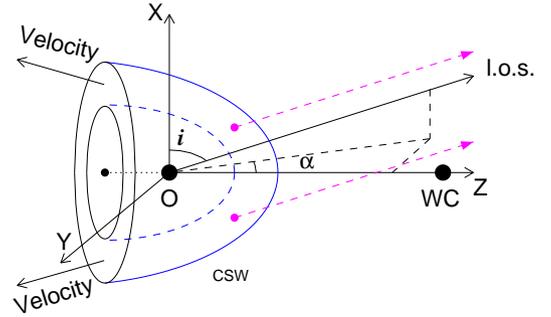}
\end{center}
\caption{ 
Schematic diagram of colliding stellar winds in a massive WC$+$O 
binary system (e.g., \WR).The wind interaction `cone' is denoted by  
CSW (the axis Z is its axis of symmetry; the axis X is perpendicular 
to the orbital plane; the axis Y completes the right-handed coordinate
system). The line-of-sight towards observer is denoted by l.o.s. and
the two related angles, $i$ (orbital inclination) and $\alpha$
(azimuthal angle) are marked as well. The arrows labelled `Velocity'
indicate the general direction of the bulk gas velocity in the
interaction region. The two dashed-line arrows (in magenta colour) 
illustrate that emission from a parcel of gas in the CSW region may
experience different wind absorption (passing closer or further from
one or both stars in the binary): this depends on the azimuthal angle
and rotational angle around the axis of symmetry (Z).
}
\label{fig:cratun}
\end{figure}

The standard physical picture of CSWs in massive binaries considers
interaction of two spherically-symmetric stellar winds, which have 
reached their terminal velocities in front of the shocks.
The interaction region has cylindrical symmetry and
two-dimensional (2D) numerical hydrodynamic models could be used for
calculating the physical parameters of the CSW structure.
Figure~\ref{fig:cratun} shows a schematic diagram of the wind 
interaction of two spherically-symmetric stellar winds in a WR$+$O 
binary. We note that here, as in our previous works, we adopt a 
convention that the O star in the binary system is located at the 
origin of the coordinate system.

As shown in the early works by \citet{lm_90}, \citet{luo_90},
\citet{stevens_92} and \citet{mzh_93}, 
the basic input parameters for the CSW hydrodynamic
model in WR$+$O binaries are the mass loss and velocity of the stellar
winds of the binary components and the binary separation.
Following the parameter notation from \citet{mzh_93}, we note that
the shape and the structure of the CSW interaction
region are defined by a dimensionless
parameter $\Lambda = (\dot{M}_{WR} V_{WR}) / (\dot{M}_{O} V_{O})$
representing the ram-pressure ratio of the stellar winds of the binary
components.

It is important to note that various mechanisms may play an important 
role for the physics of the CSW region. For example, CSW shocks could
be adiabatic or radiative, partial electron heating might occur behind
strong shocks, non-equilibrium ionization (NEI) effects could affect
the X-ray emission from the CSW region. The importance of each
physical mechanism could be estimated based on some dimensionless 
parameters: 
(a) parameters 
$\chi$ (see eq. 8 in \citealt{stevens_92}) and 
$\Gamma_{ff}$ (see eq.9 in \citealt{mzh_93})
show whether or not the shock plasma is adiabatic;
(b) parameter 
$\Gamma_{eq}$ (see eq. 1 in  \citealt{zhsk_00})
determines whether the difference of 
electron and ion temperatures is important;
(c) parameter  
$\Gamma_{NEI}$ (see eq. 1 in \citealt{zh_07})
estimates the significance of the NEI 
effects.

\begin{figure}
\begin{center}
\includegraphics[width=\columnwidth]{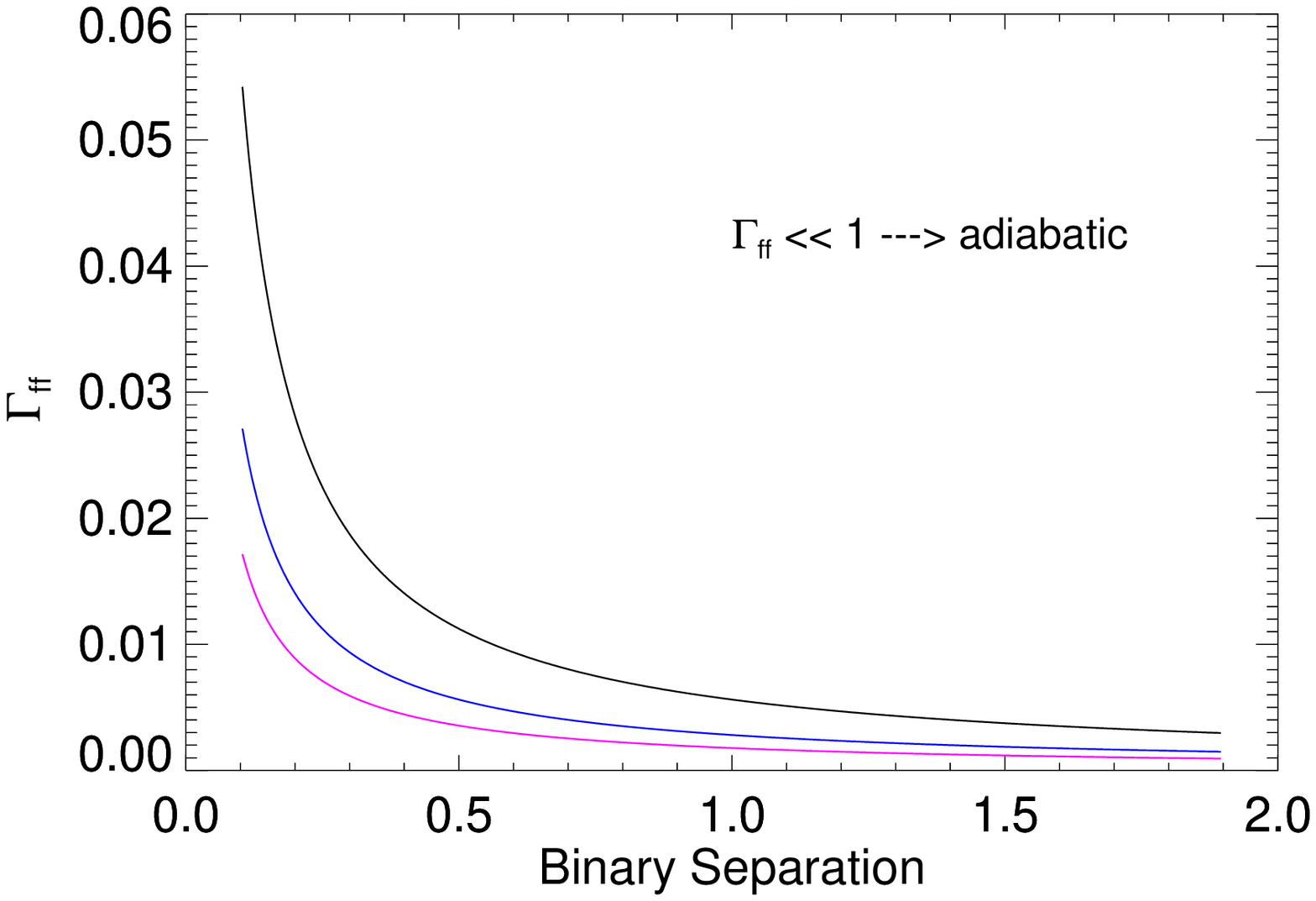}
\includegraphics[width=\columnwidth]{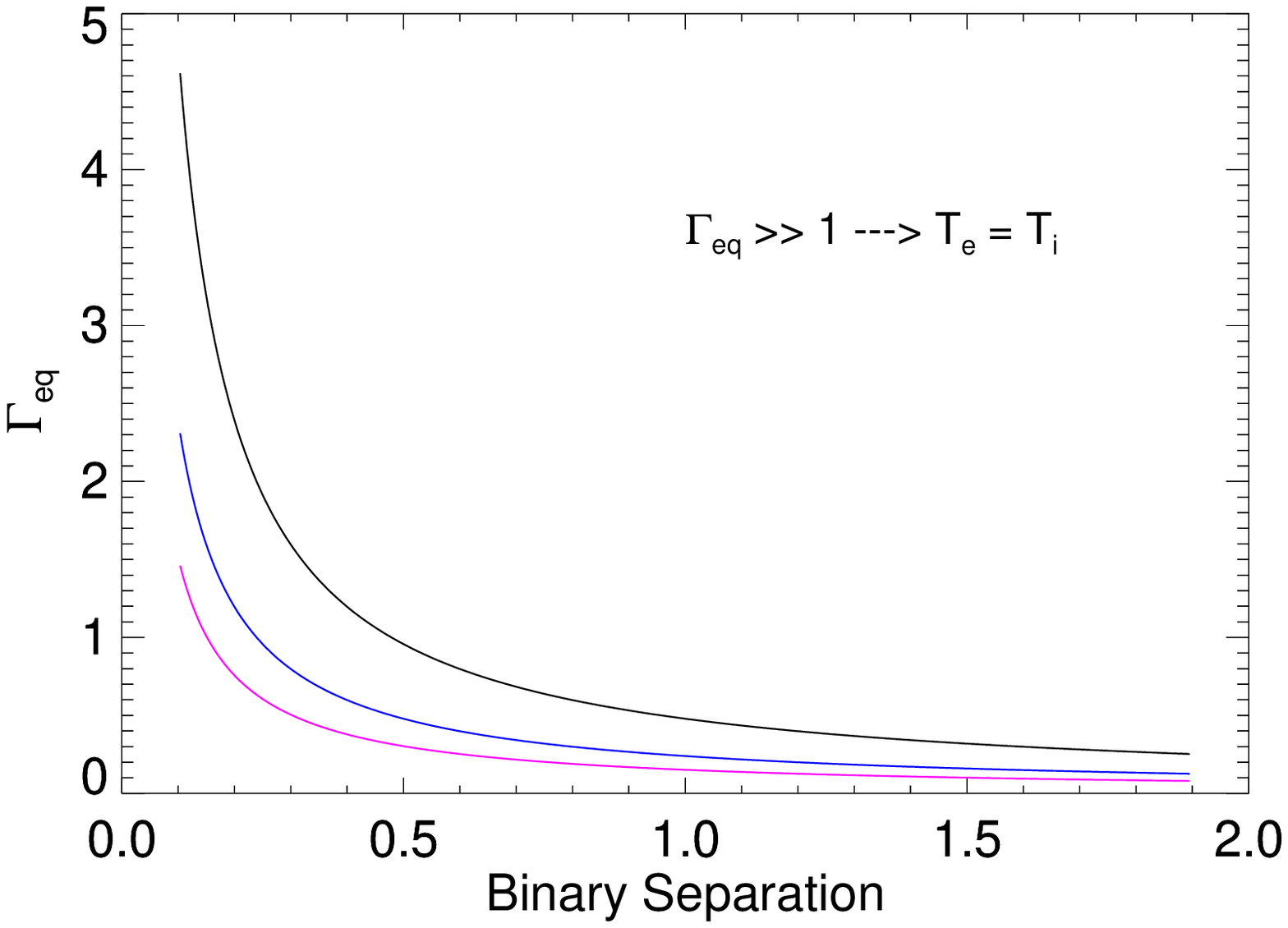}
\includegraphics[width=\columnwidth]{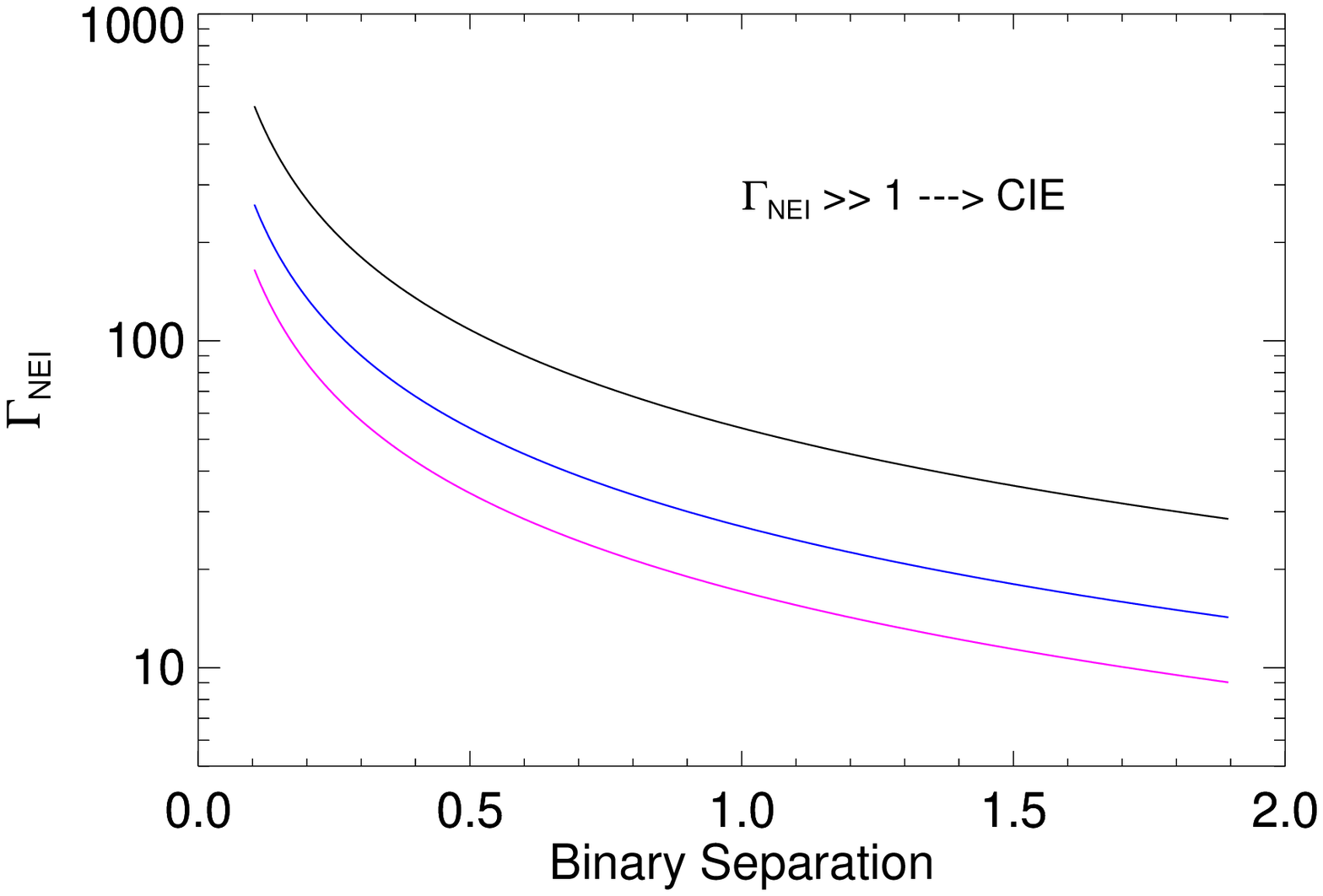}
\end{center}
\caption{The dependence of dimensionless CSW parameters $\Gamma_{ff}$,
$\Gamma_{eq}$ and $\Gamma_{NEI}$ on the range of binary separation
over the entire orbit of \WRE. For comparison, the cases with reduced
mass-loss rates of 0.25 and 0.1 their nominal values are shown in blue
and magenta colour, respectively.
Binary separation is in units of the semi-major axis.
}
\label{fig:gamma}
\end{figure}

Based on the nominal values of the stellar wind parameters and binary 
separation (Section~\ref{sec:star}), Figure~\ref{fig:gamma} shows the
corresponding values of these dimensionless parameters for the entire
orbit of \WR (i.e., the range of binary separation 
$0.1036 - 1.8964$ in units of semi-major axis).
We see that CSW models that consider adiabatic shocks are suitable for
the case of \WRE. The shock-heated plasma should have different
electron and ion temperatures (T$_e \neq$ T$_i$). And, the NEI effects 
are not important that is the X-ray emission originates from plasma in
collisional ionization equilibrium (CIE).

For modelling all X-ray spectra of \WR in this study,
these estimates justify the 
use of our CSW \xspec model that takes into account 
the different electron and ion temperature behind the shocks.
All details of the model are found in \citet{zhsk_00} but we
nevertheless list the basic physical features of the model.
Namely, along the entire shock surfaces of the CSW region the ratio of
the local post-shock {\it electron} temperature to the mean post-shock 
plasma temperature is defined by a single parameter ($\beta$). The
evolution of the electron temperature is followed along each
streamline behind the shock surfaces. No thermal broadening is
considered. The latter has no effect for undispersed X-ray spectra but
might have some effect for high-resolution grating spectra. However,
anticipating some basic results from the current analysis 
(see Section \ref{sec:fit_results}; Figs.~\ref{fig:lines_meg},
\ref{fig:lines_heg}, we are confident to assume that considering the 
effects of thermal broadening could be left beyond the scope of this 
study.

For modelling the X-ray spectra with high spectral resolution, we
used the \xspec model that explicitly takes into account the line
broadening (bulk gas velocities) from the hydrodynamic CSW model
\citep{zhp_10}. A new feature was also added in this \xspec model,
namely, it now considers the specific stellar wind (WR and/or O-star) 
absorption along the line of sight to the observer for each parcel of 
gas in the CSW region. This is done in a `cold'-wind approximation.

It is important to note that our \xspec models can take into account
the different chemical composition of the WR and O-star wind  in
modelling the X-ray emission from the CSW region in \WRE. This is
because they are based on the 2D numerical hydrodynamic model of
CSW by \citet{lm_90} (see also \citealt{mzh_93}), which provides an
exact solution to all discontinuity surfaces (the two shocks and the 
contact discontinuity) of the CSW region. This means that the X-ray
emission from the shocked WR and O-wind plasma can be modelled 
separately.

Finally, for clarity we provide the details of our fitting
procedure as given in section 4.1 of \citet{zh_17}.
"Our entire fitting procedure is threefold: 
(a) given the stellar wind and binary parameters, the hydrodynamic
model provides the physical parameters of the CSW region;
(b) based on these results, we prepare the input quantities 
(e.g., distribution of temperature, emission measure, ionization age,
bulk gas velocity) 
for the {\it spectral} model in \xspecE;
(c) the CSW \xspec  model fits the observed X-ray spectrum.
Note that in \xspec we can fit for the X-ray absorption, chemical
abundances and the model normalization parameter.
If adjustments of other physical parameters are required, all three
steps should be repeated: i.e., our fitting procedure is an iterative
process.
It is important to keep in mind that the normalization parameter
($norm$) of the CSW model in \xspec is a dimensionless
quantity that gives the ratio of observed to theoretical fluxes. 
Thus, the entire fitting procedure is aimed at getting a value of 
$norm = 1.0$, which indicates a perfect match between the 
observed count rate and that predicted by the model 
($norm < 1.0$ indicates a theoretical flux higher than that required 
by the observations and the opposite is valid for $norm > 1.0$)."

Since we are dealing with X-ray emission from thermal plasma, $norm =
1$ (or $norm < 1.0$ ; $norm > 1.0$) shows whether the theoretical 
model predicts an amount of emission measure exactly as needed (or 
higher/smaller then that) to explain the observed X-ray emission.

\begin{table}

\caption{\WR Spectral Fit Results (abundances)
\label{tab:fits}}
\begin{center}
\begin{tabular}{lll}
\hline
\multicolumn{1}{c}{ } &
\multicolumn{1}{c}{C / He = 0.1}  &
\multicolumn{1}{c}{C / He = 0.4}  \\
\hline
Ne &  $0.75\pm0.02$ &  $1.10\pm0.06$ \\
Mg &  $0.29\pm0.03$ &  $0.44\pm0.03$ \\
Si &  $0.93\pm0.04$ &  $1.37\pm0.07$ \\
S  &  $1.67\pm0.08$ &  $2.43\pm0.11$ \\
Ar &  $1.91\pm0.11$ &  $2.50\pm0.35$ \\
Ca &  $0.52\pm0.35$ &  $0.55\pm0.51$ \\
Fe &  $1.14\pm0.05$ &  $1.61\pm0.15$ \\
\hline

\end{tabular}
\end{center}

{\it Note}. Abundance values derived from the simultaneous fits to the
\ChandraE-MEG spectra by making use of the CSW model that takes into
account different electron and ion temperatures:
$\beta = T_e / T$, $T_e$ is the electron temperature and $T$ is the
mean plasma temperature. Given are the mean value for
each element and its standard deviation for the entire set of 12
values of $\beta \in [0.001, 1] $ (see Section~\ref{sec:fit_results}).

\end{table}

\subsection{CSW model spectral results}
\label{sec:fit_results}
In all the spectral fits, 
we explored a range of values for the partial heating of the
electrons at the shock fronts:
$\beta = 
    [0.001, 0.05, 0.1, 0.2, 0.3, 0.4, 0.5, 0.6, 0.7, 0.8, 0.9, 1]$;
$\beta = T_e / T$, $T_e$ is the
electron temperature and $T$ is the mean plasma temperature
($\beta < 1 \rightarrow$ 2-T plasma, different
electron and ion temperatures; $\beta = 1 \rightarrow$ 1-T plasma,
equal ion and electron temperatures).

We considered two components of the X-ray
absorption: (a) due to the interstellar matter (ISM); (b) due to the
stellar wind(s) in \WRE. 
For the ISM absorption component (kept fixed in the fits), we adopted
a value of the foreground column density corresponding to the optical 
extinction $A_v = 2.50$~mag ($A_V = A_v / 1.11$ and table 28 in 
\citealt{vdh_01}) and the \citet{go_75} conversion 
N$_H = 2.22\times10^{21}$A$_{\mbox{V}}$~cm$^{-2}$.
The stellar wind absorption shared the same abundances that were
adopted for the X-ray emitting plasma in the CSW region.

We note that the chemical abundances of the shocked 
O-star wind were solar \citep{ag_89}. For the shocked WR-star wind, we 
explored two values for the carbon abundance of   
C/He = 0.1 and 0.4 by number (bracketing the possible range as
deduced from observations of WC stars, 
see \citealt{ee_wi_92}, \citealt{dessart_00}
also including the value typically adopted in the theoretical models
of optical spectra of WC stars, e.g. see table 2 in
\citealt{sander_12}), 
while the other elements had their 
values typical for the WC stars (by number) as from \citet{vdh_86}:
H = 0.0, He = 1.0, C = 0.1 (0.4), N = 0.0, O = 0.194, 
Ne = $1.86\times10^{-2}$, Mg = $2.72\times10^{-3}$, 
Si = $6.84\times10^{-4}$, S = $1.52\times10^{-4}$, 
Ar = $2\times10^{-5}$, Ca = $2\times10^{-5}$, 
Fe = $3.82\times10^{-4}$.
Ar and Ca are not present in the \citet{vdh_86} abundance set, so, we
adopted for each of them a fiducial value of $2\times10^{-5}$.
It is worth mentioning that the contribution of the shocked O-star
wind to the total observed X-ray emission (flux) of \WR is not higher
than 5-6\%.
So, the Ne, Mg, Si,
S, Ar, Ca, Fe abundances of the shocked WR plasma were allowed to vary 
to improve the quality of the fits.
Because chemical abundances are not well constrained from undispersed
X-ray spectra, we went through the following steps.

1) For each value of parameter $\beta$, the X-ray spectra with best 
photon statistics (\ChandraE-MEG, re-binned to have a minimum of 20 
counts per bin) were fitted simultaneously to estimate the abundances.
The \xspec model {\it gsmooth} was adopted for the line broadening to 
minimize the amount of CPU time. So, in \xspec terms  the fitted 
two-absorption model reads:
$Spec = wabs(ISM) * wabs(wind) * gsmooth(csw)$, where $csw$ is our CSW
model with line-broadening switched off
. As seen from Table~\ref{tab:fits},
there is no big scatter between the abundance values for a specific
chemical element.
We note that the quality of the fits was good in a formal statistical
sense for all the values of parameter $\beta$: 
for $\beta \in [0.001, 1] \rightarrow \chi^2 \in [4636,
4938]$ (C/He = 0.1) and $\chi^2 \in [4525, 4952]$ (C/He = 0.4)
with 5070 degrees of freedom (dof).

2) The derived abundances for each $\beta$ were used (kept
fixed) in the  fits of the \rxte spectra adopting the same model as in
the first step but with no line broadening (no $gsmooth$ component). 
In these fits, free parameters were the X-ray absorption and the 
normalization parameter {\it norm}. 

\begin{figure*}
\begin{center}
\includegraphics[width=\columnwidth]{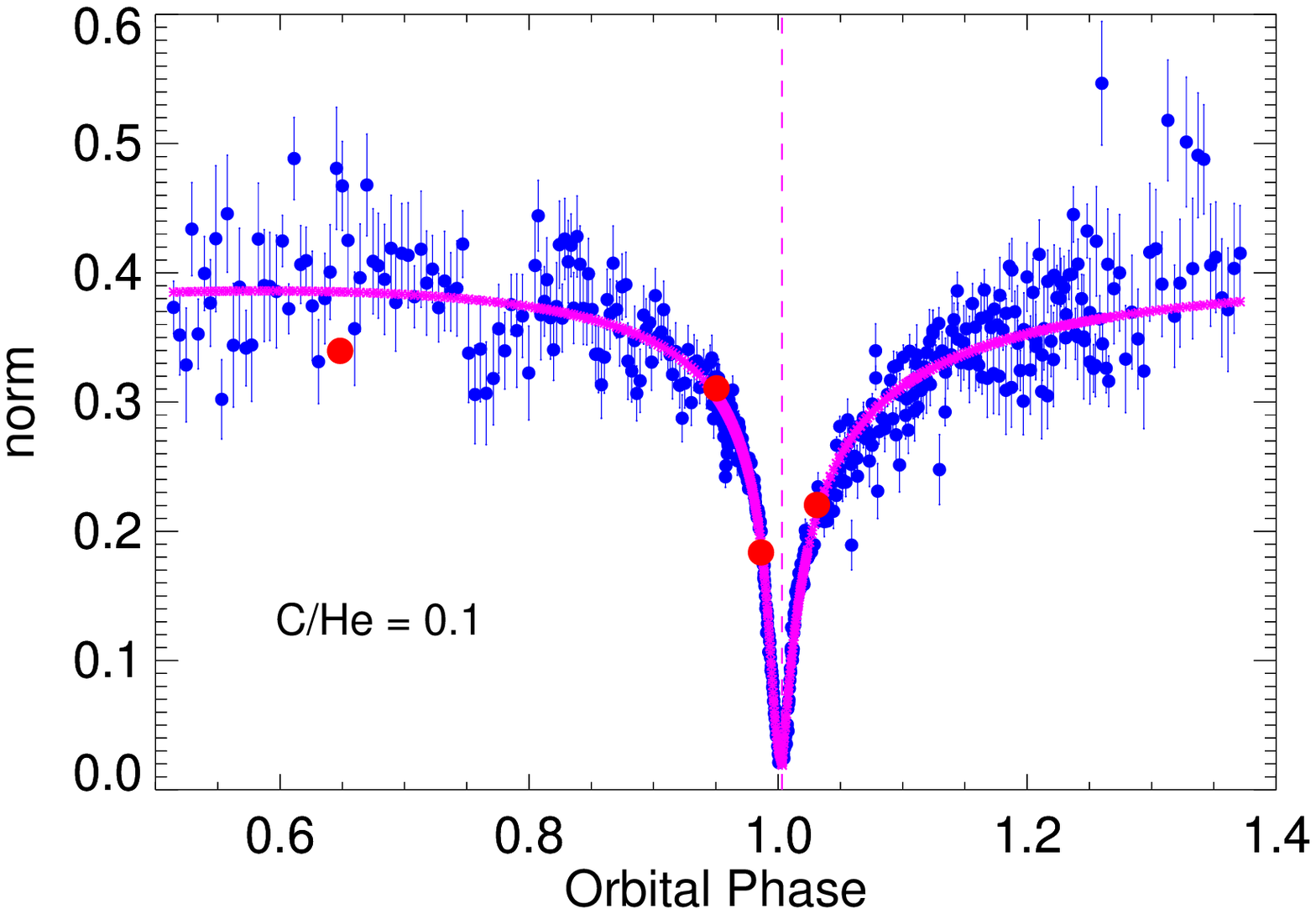}
\includegraphics[width=\columnwidth]{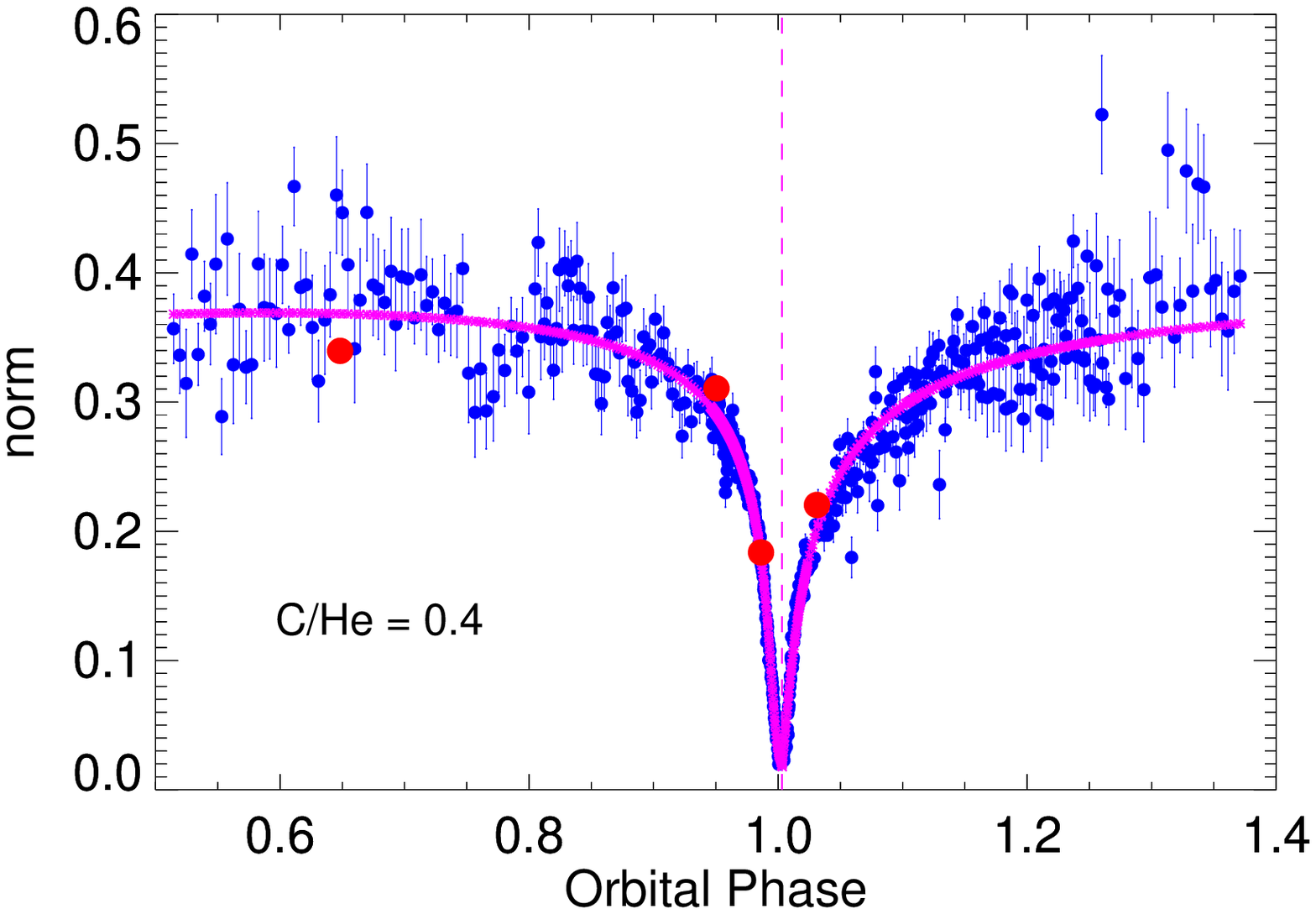}
\includegraphics[width=\columnwidth]{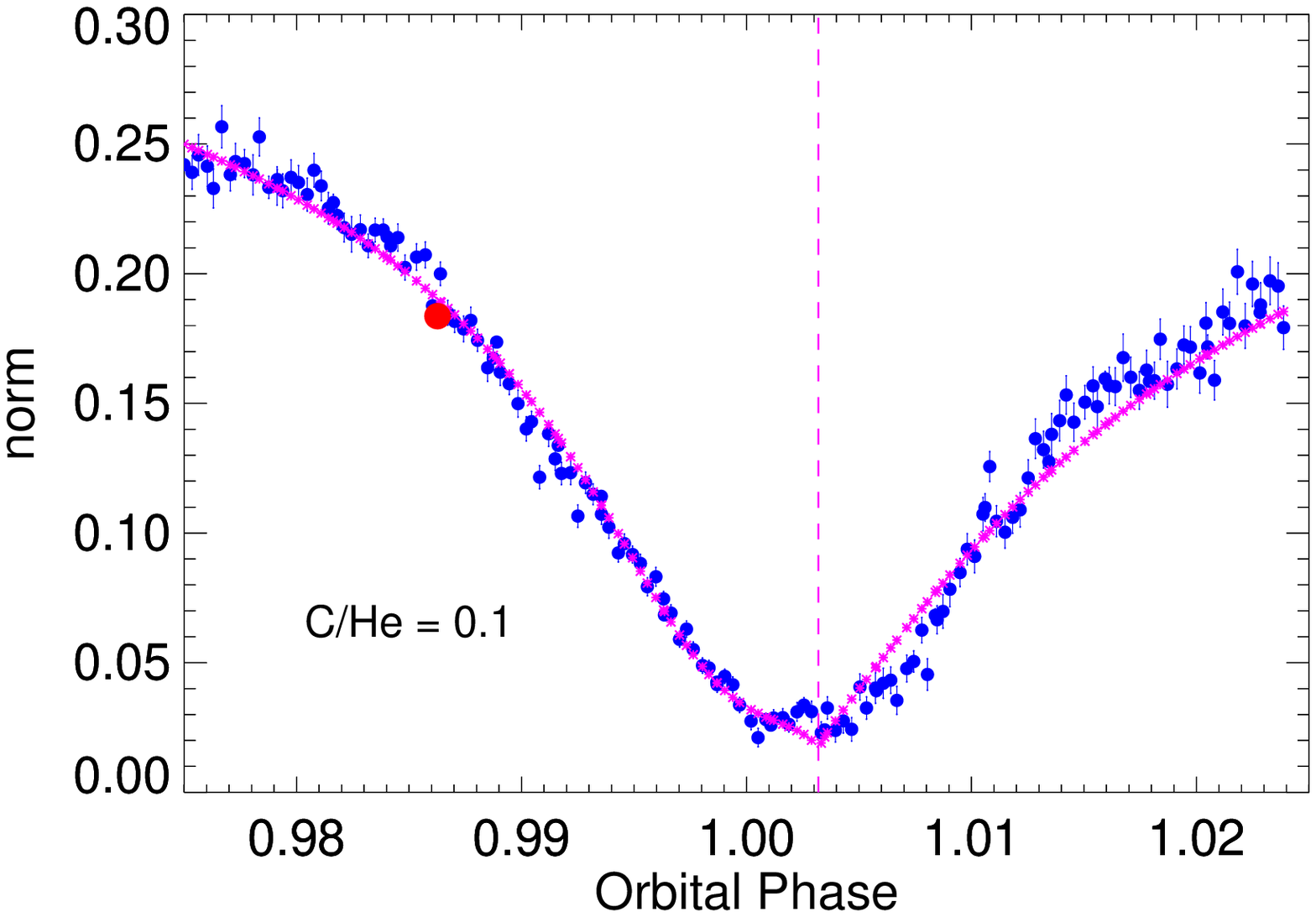}
\includegraphics[width=\columnwidth]{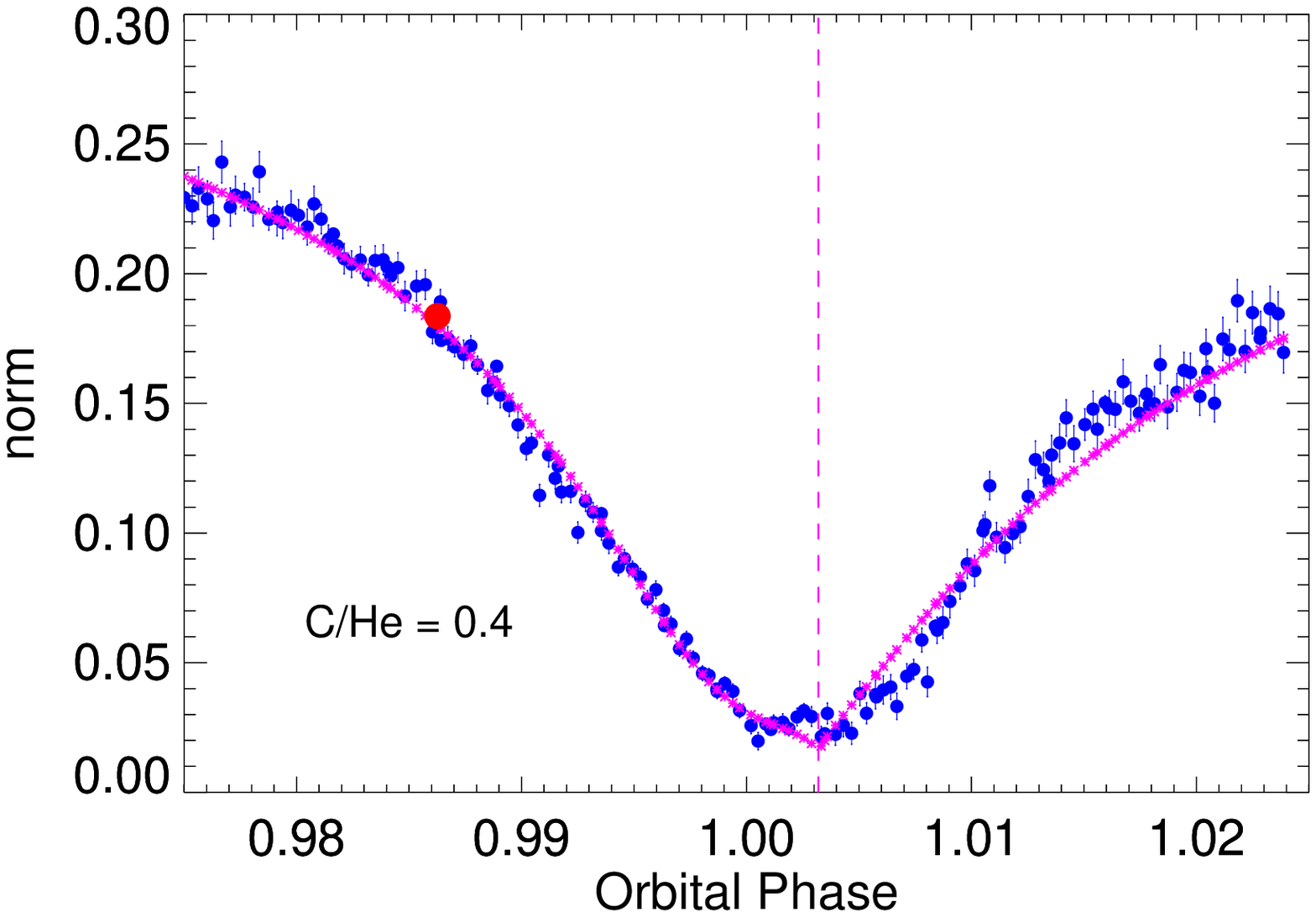}
\end{center}
\caption{The values of the normalization parameter ($norm$) of the CSW
\xspec model as derived from the fit to the \rxte (blue dots)
and \Chandra (red filled circles) spectra of \WR for the case of
different electron and ion temperature ($\beta = 0.1$) and different 
chemical composition (C/He = 0.1; 0.4). The vertical dashed line 
denotes the orbital phase of WR-star in front.
The two-component fit to the normalization parameter (see text for
details) is overlaid in magenta colour.
}
\label{fig:norm}
\end{figure*}

X-ray absorption is found to vary over the binary orbit. And,
we note that the most important result from these fits is that a
considerable decrease is found for the $norm$ parameter at orbital
phases near periastron. This is indicative of changing amount of
emission measure over the binary orbit.
We recall that the emission measure (EM) in the CSW region that 
results from interaction of spherically-symmetric stellar winds is 
proportional to the square of the stellar wind mass loss ($\dot{M}$) 
and is reversely proportional to the binary separation:
EM $\propto n^2 V$, $n$ is the number density, $V$ is
the volume; $n \propto \dot{M}/a^2$ and $V \propto a^3$,
therefore EM $\propto \dot{M}^2/a$. Since the $1/a$-dependence of 
emission measure on the binary separation is accounted for in the 
hydrodynamic simulations, the fit results are thus suggestive that 
mass loss varies over the binary orbit.
A closer look at derived values of the $norm$ parameter near
periastron (e.g., bottom panels in Fig.~\ref{fig:norm}) reveals that 
EM decreases gradually when approaching periastron (phase $= 0$, i.e.
phase $= 1$) but it does not increase after that. Instead, it
levels off and starts increasing only after the orbital phase the WR
stellar component is in front.

We thus propose to represent the total variation of EM (the $norm$
parameter) by two terms. The first one is responsible for the 
suggestive variation of the mass loss with binary separation, so, it 
is symmetric with respect to orbital phase $= 0$ (periastron). The
second one is symmetric with respect to the orbital phase the WR star
is in front (phase $= 0.00319$; Section~\ref{sec:star}).
We note that the basic `requirement' imposed on the specific function 
describing each term is to gradually decrease at the corresponding 
phase of symmetry and level off at phases far from it. We underline
that the choice of functions in mathematical sense is {\it not}
important as long as they have the mentioned behaviour. For the first
term, $\dot{M} (a)$,  we assume that binary separation ($a$) is the 
independent variable.  For the second term, $Scl (\alpha)$, we assume 
that azimuthal angle ($\alpha$; Fig.~\ref{fig:cratun}) is the 
independent variable.
We recall that the binary separation is in the range 
$a \in [1 - e, 1 + e]$
(in units of the semi-major axis) , 
where $e = 0.8964$ (Section~\ref{sec:star}) is 
the orbital eccentricity. Also, the azimuthal angle is in range
$\alpha \in [0, 180]$~degrees due to suggested symmetry, that is
the $Scl (\alpha)$ values for $\alpha$ and $\alpha = 360 - \alpha$  
are the same. We have to keep in mind that given the binary ephemeris
for each orbital phase we know the binary separation and the azimuthal
angle, that is there is a unique correspondence between the latter two
quantities. We thus write:

$$
\dot{M}(a) = \dot{M}_1 + (\dot{M}_2 - \dot{M}_1) \times 
\left[1 - \left(\frac{1 - e}{a}\right)^{s_1}\right]^{s_2}
$$

\[
Scl({\alpha}) = \left\{
   \begin{array}{ll}
   1 & if \hspace{0.1cm} \alpha \geq \alpha_0 \\
   b_1 + (1 - b_1) \times 
   \tanh\left[\left(\frac{\alpha}{\alpha_0 - \alpha}\right)^{b_2}\right]
   & if  \hspace{0.1cm} \alpha < \alpha_0 
   \end{array}
\right.
\]

\begin{equation}
\hspace{1.5cm}  norm = [\dot{M}(a)]^2 \times Scl(\alpha)
\label{eqn:norm}
\end{equation}
where $\dot{M}_1, \dot{M}_2, s1, s2, b_1, \alpha_0, b_2$ are free 
parameters in the fits.

For each case (value) of $\beta$,
we used eq.~\ref{eqn:norm} to fit the normalization parameter values 
derived from the spectral fits. We found that eq.~\ref{eqn:norm} is a 
good representation of the changes of this parameter over the orbital
period: an example of the corresponding results is shown in 
Fig.~\ref{fig:norm}. The two terms $\dot{M} (a)$ and $Scl (\alpha)$
derived for 
the `best' cases of $\beta < 0.4$ (see below) are shown in 
Fig.~\ref{fig:norm_fit} and the corresponding coefficients in
eq.~\ref{eqn:norm} are given in Table~\ref{tab:norm}.

\begin{table}

\caption{Coefficients in eq.~\ref{eqn:norm}
\label{tab:norm}}
\begin{center}
\begin{tabular}{lll}
\hline
\multicolumn{1}{c}{ } &
\multicolumn{1}{c}{C / He = 0.1}  &
\multicolumn{1}{c}{C / He = 0.4}  \\
\hline
$\dot{M}_1$ &  $0.243\pm0.001$ &  $0.236\pm0.001$ \\
$\dot{M}_2$ &  $0.689\pm0.010$ &  $0.677\pm0.010$ \\
$s_1$       &  $0.667\pm0.027$ &  $0.656\pm0.026$ \\
$s_2$       &  $1.003\pm0.015$ &  $1.008\pm0.015$ \\
  &  & \\
$b_1$       &  $0.199\pm0.002$ &  $0.200\pm0.002$ \\
$b_2$       &  $0.644\pm0.006$ &  $0.650\pm0.007$ \\
$\alpha_0$  &  $177.5\pm0.739$ &  $178.5\pm0.572$ \\
\hline

\end{tabular}
\end{center}

{\it Note}. Values of coefficients in eq.~\ref{eqn:norm} derived from 
the fits to the normalization parameter ($norm$) as obtained from the 
fits to the \rxte spectra of \WRE. Given are the mean value for each 
coefficient and its standard deviation for the set' of 
$\beta < 0.4$ values (see Section~\ref{sec:fit_results}).

\end{table}

\begin{figure*}
\begin{center}
\includegraphics[width=\columnwidth]{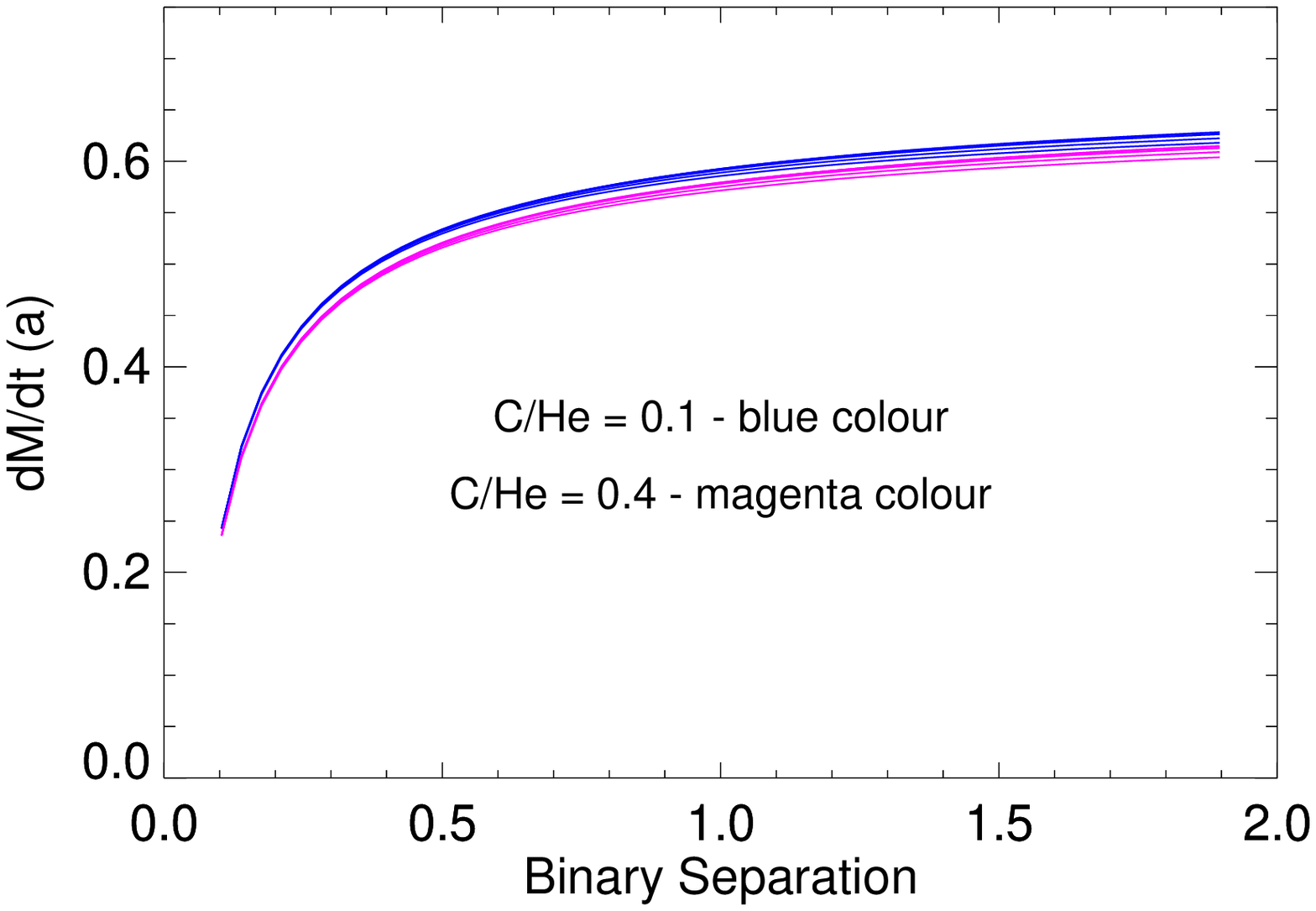}
\includegraphics[width=\columnwidth]{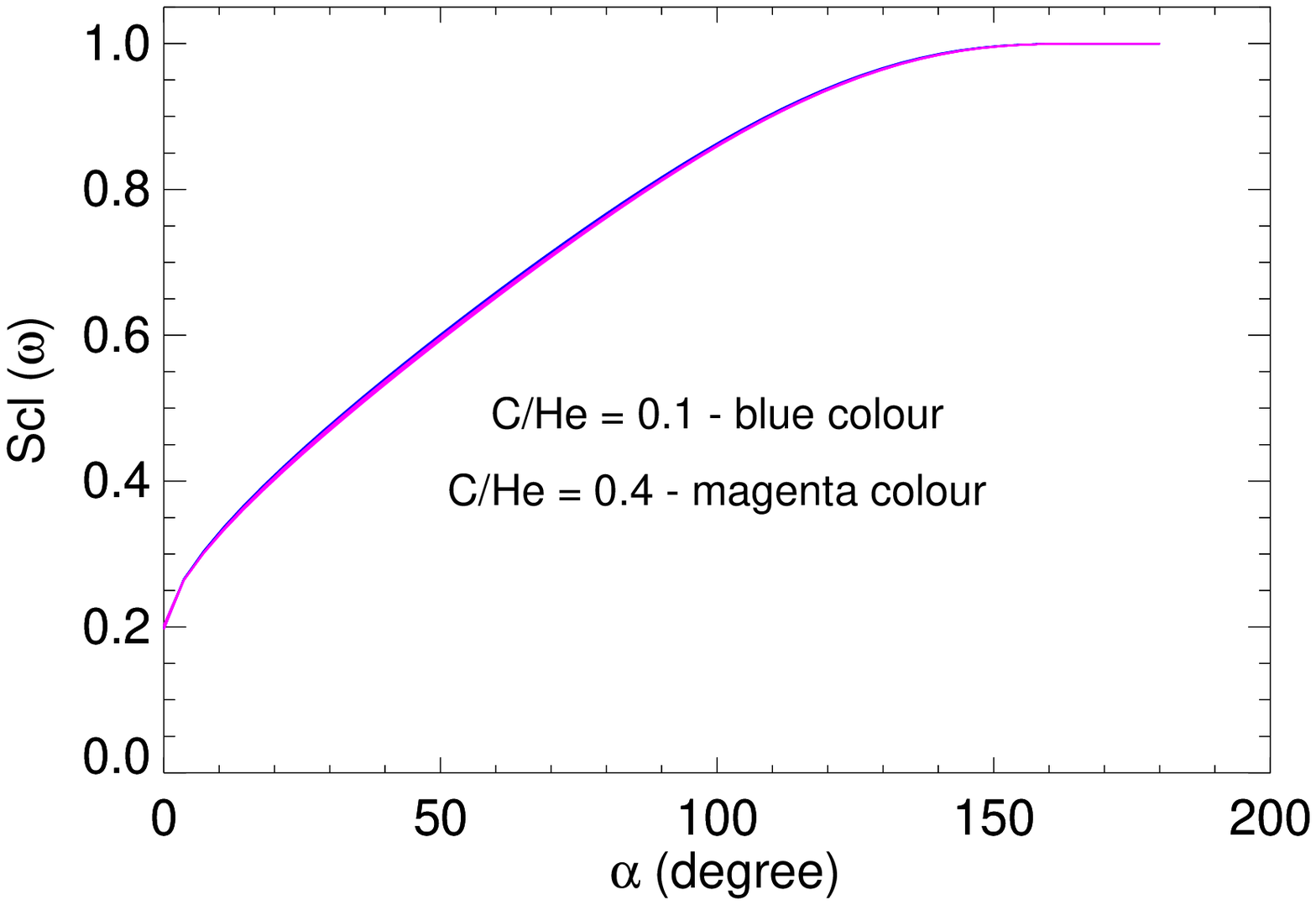}
\end{center}
\caption{Results from the use of eq.~\ref{eqn:norm} for fitting
the values of the normalization parameter ($norm$) over the orbital
period for the `best' set of $\beta < 0.4$ values
(Section~\ref{sec:fit_results}). The first term $\dot{M} (a)$ and the
second term $Scl (\alpha)$ is shown int the left and right panel,
respectively.
Binary separation is in units of the semi-major axis.
}
\label{fig:norm_fit}
\end{figure*}

\begin{figure*}
\begin{center}
\includegraphics[width=\columnwidth]{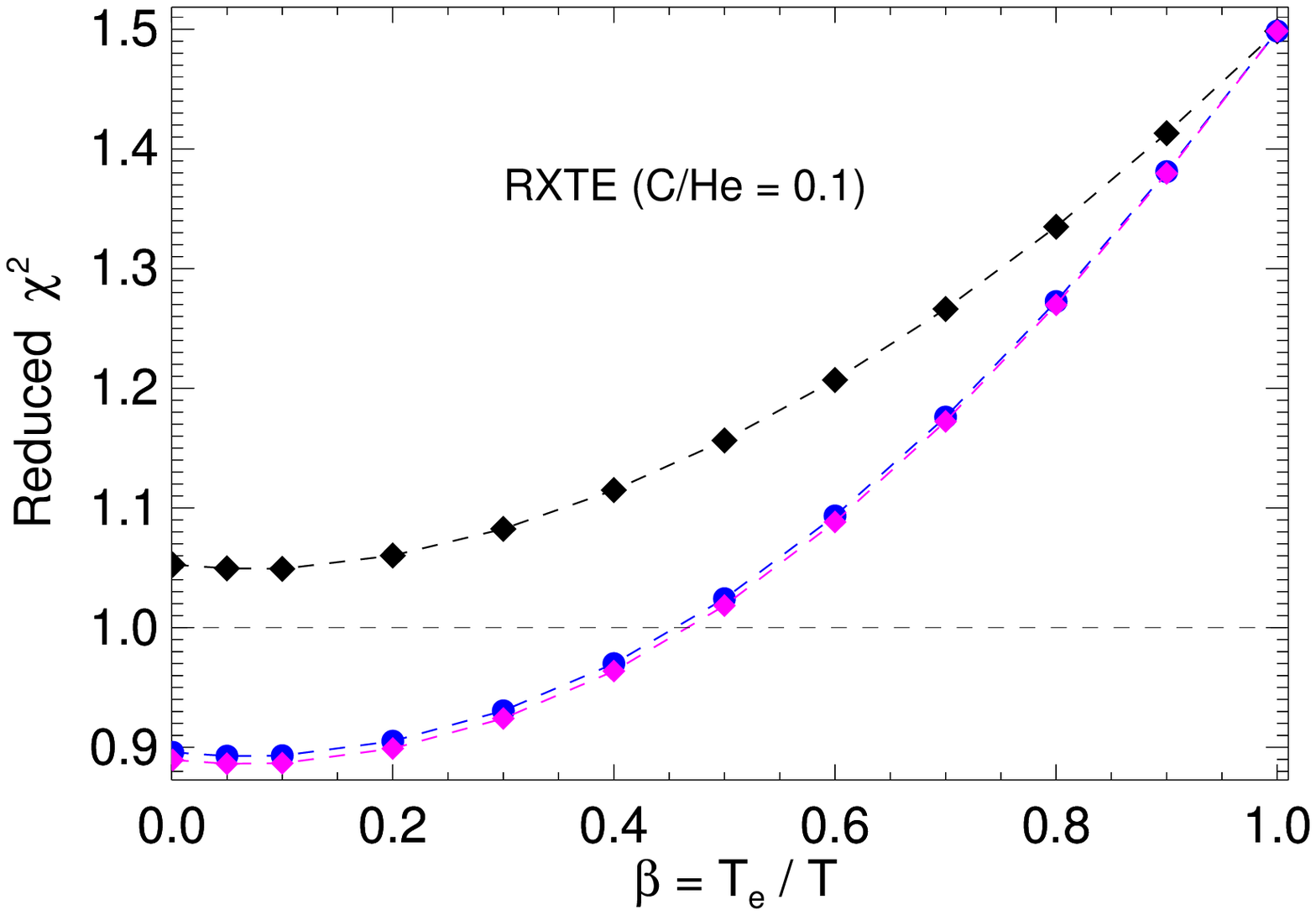}
\includegraphics[width=\columnwidth]{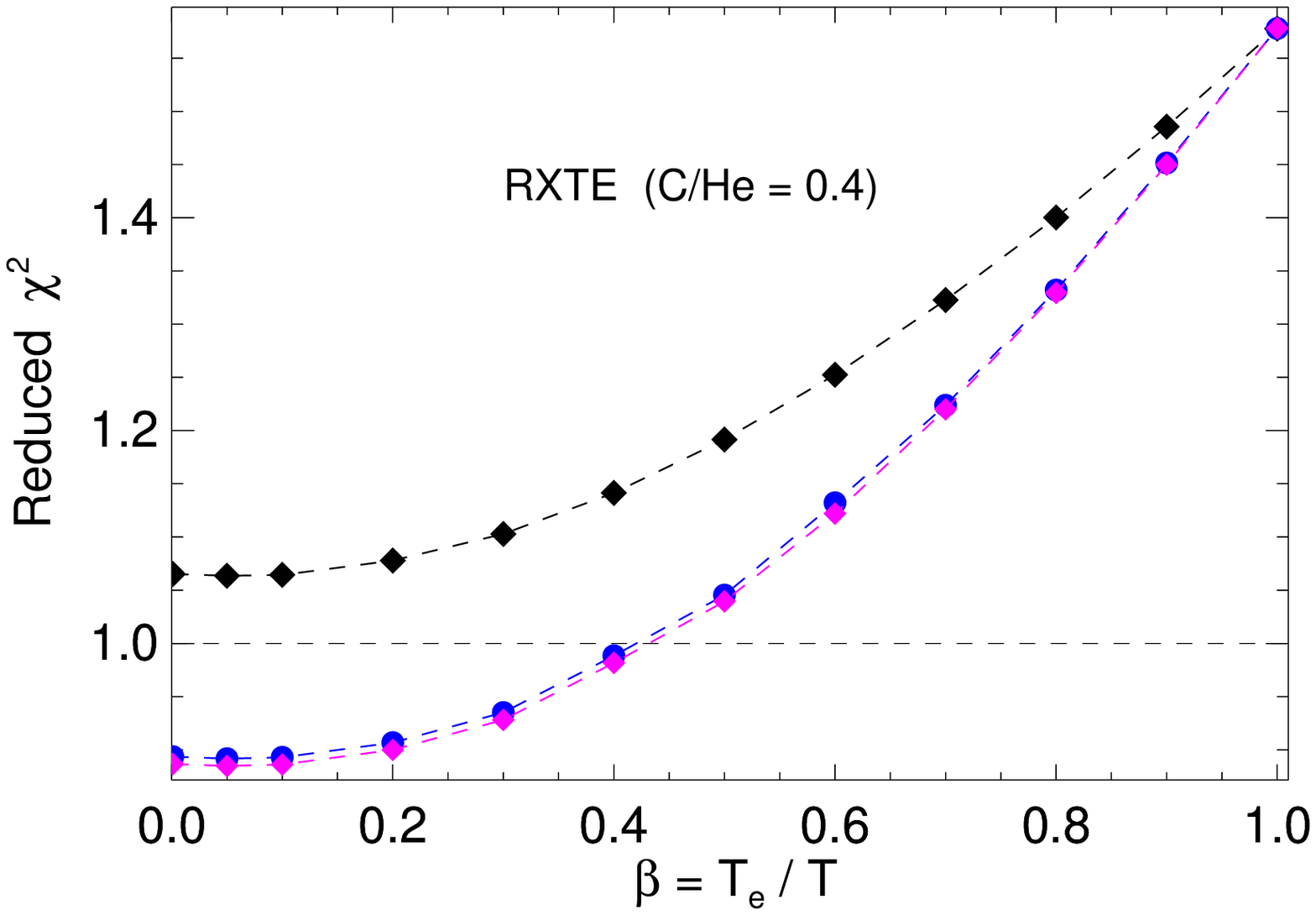}
\end{center}
\caption{The total reduced $\chi^2$ (dof $= 22080$) for the entire set
of 552 \rxte spectra. The top curve (in black) shows the results
from the CSW spectral fits with the nominal stellar-wind parameters.
The lower two curves presents the results with reduced mass-loss rates
as described by terms $\dot{M} (a)$ (in blue colour) and $Scl (\alpha)$ 
(in magenta colour) of eq.~\ref{eqn:norm}, respectively. The $\chi^2$
values are provided in Table~\ref{tab:chi2}.
}
\label{fig:chi2}
\end{figure*}

As described in Section~\ref{sec:csw}, the entire spectral fitting
procedure is aimed at getting a value of $norm = 1$. So, we made two
more general iterations, adopting the reduced mass-lass rates
corresponding to the derived values for each term in
eq.~\ref{eqn:norm}. Namely, we had to re-run the hydrodynamic
simulations, derive the corresponding distribution of the physical
parameters in the CSW region and re-do the spectral fits. We note that
all this resulted not only in values of $norm \approx 1$ but in a
better quality of the spectral fits. As seen from Fig.~\ref{fig:chi2},
the basic improvement is due to the reduced mass-loss rates ($\dot{M}
(a)$) and the quality of the fits does not get better by considering
the effect of the second term ($Scl (\alpha)$) of eq.~\ref{eqn:norm}.

\begin{figure*}
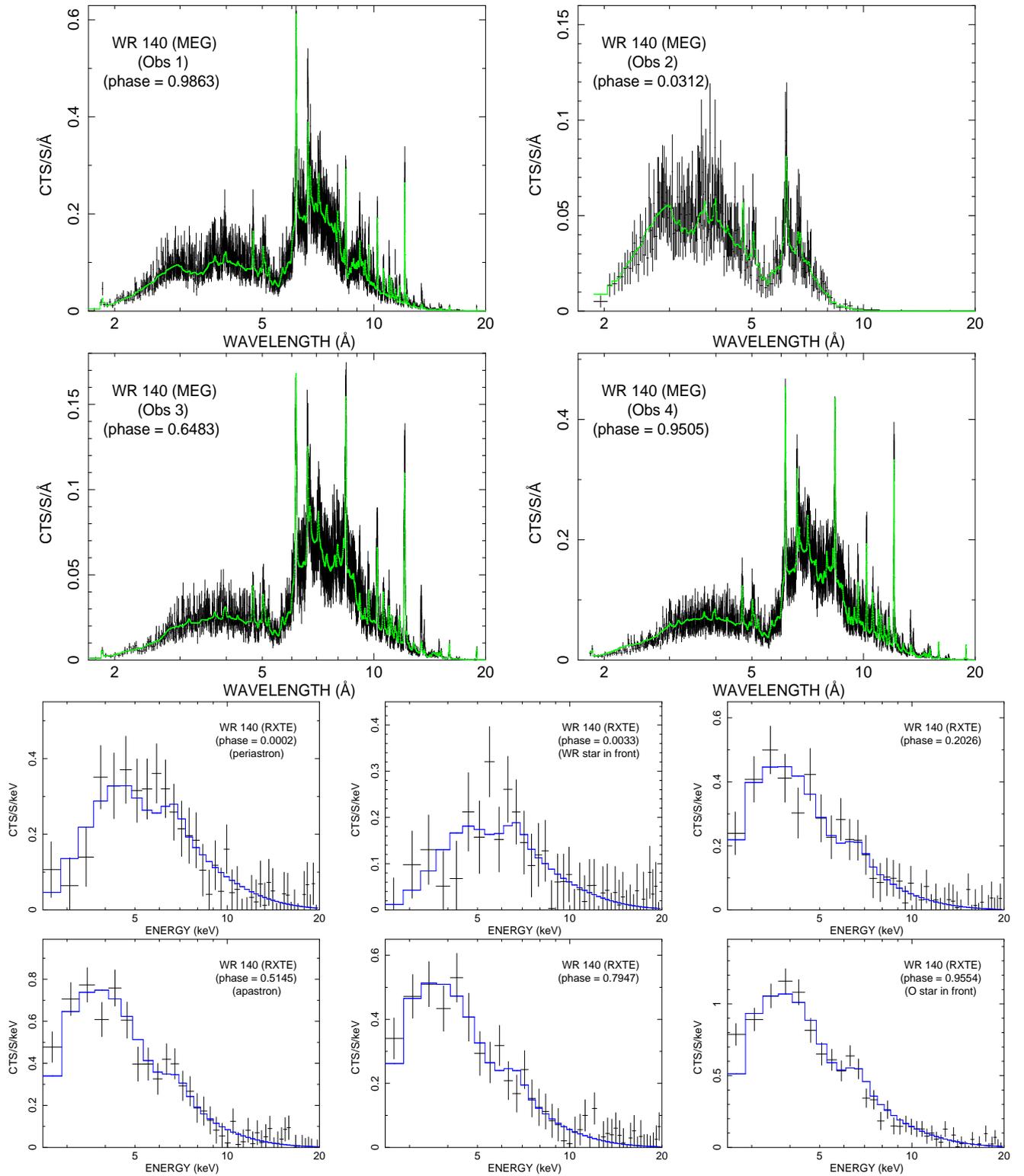

\begin{center}
\includegraphics[width=2.357in, height=3.3in,angle=-90]{fig7a.eps}
\includegraphics[width=2.357in, height=3.3in,angle=-90]{fig7b.eps}
\includegraphics[width=2.357in, height=3.3in,angle=-90]{fig7c.eps}
\includegraphics[width=2.357in, height=3.3in,angle=-90]{fig7d.eps}
\includegraphics[width=1.6in,height=2.3in,angle=-90]{fig7e.eps}
\includegraphics[width=1.6in,height=2.3in,angle=-90]{fig7f.eps}
\includegraphics[width=1.6in,height=2.3in,angle=-90]{fig7g.eps}
\includegraphics[width=1.6in,height=2.3in,angle=-90]{fig7h.eps}
\includegraphics[width=1.6in,height=2.3in,angle=-90]{fig7i.eps}
\includegraphics[width=1.6in,height=2.3in,angle=-90]{fig7j.eps}
\end{center}
\caption{
Background-subtracted spectra of \WR and the CSW model ($\beta = 0.1$;
C/He = 0.4) fit.
The \Chandra HETG-MEG spectra were re-binned to have a minimum of 20
counts per bin, while the \rxte spectra retained their original
binning.
A label in each panel provides the value of the orbital phase for the
corresponding observation.
}
\label{fig:meg_rxte}
\end{figure*}

Also, it seems conclusive that two-temperature plasma with
$\beta < 0.4$ (i.e. partially heated electrons at the shock fronts) 
is needed in the CSW region of \WR to successfully 
explain the X-ray emission from this massive binary.
In general, we could try to define the `best value' of $\beta$, that
is the one which provides the best match to the observed spectra. But,
approaching this in a standard statistical way shows that the goodness 
of the fit is equal to 1 for all the cases with $\beta < 0.4$, which
means that all these fits are equally good. Thus,
we do not think that a more definite conclusion could be drawn rather
than the one that two-temperature plasma with $\beta < 0.4$.is present
in the CSW region of \WRE.
An illustration of the CSW spectral fits is shown in
Fig.~\ref{fig:meg_rxte}.

It is worth noting that one of the basic effects of the partial
heating of electrons in shock fronts ($\beta < 1$) is that the
electron temperature is lower (not higher) than the local 
mean plasma temperature. Therefore, the resulting X-ray emission of
the shocked plasma will be `cooler' in this case (two-temperature
plasma).
This explains why the quality of the fits to the \rxte spectra of \WR 
improves when the case of two-temperature plasma is considered. 
Namely, a closer look at the \rxte spectra established that some
excess in the {\it X-ray model emission} starts to emerge at high 
energies when $\beta$ increases and approaches unity (i.e. for equal 
ion and electron temperatures).

However, we have to keep in mind that low-resolution spectra (\rxteE)
are lacking some very important details about the gas kinematics of
the emitting X-ray plasma. Therefore, modelling the high-resolution
X-ray spectra of \WR in detail is a natural step to fill this gap.

\begin{figure*}
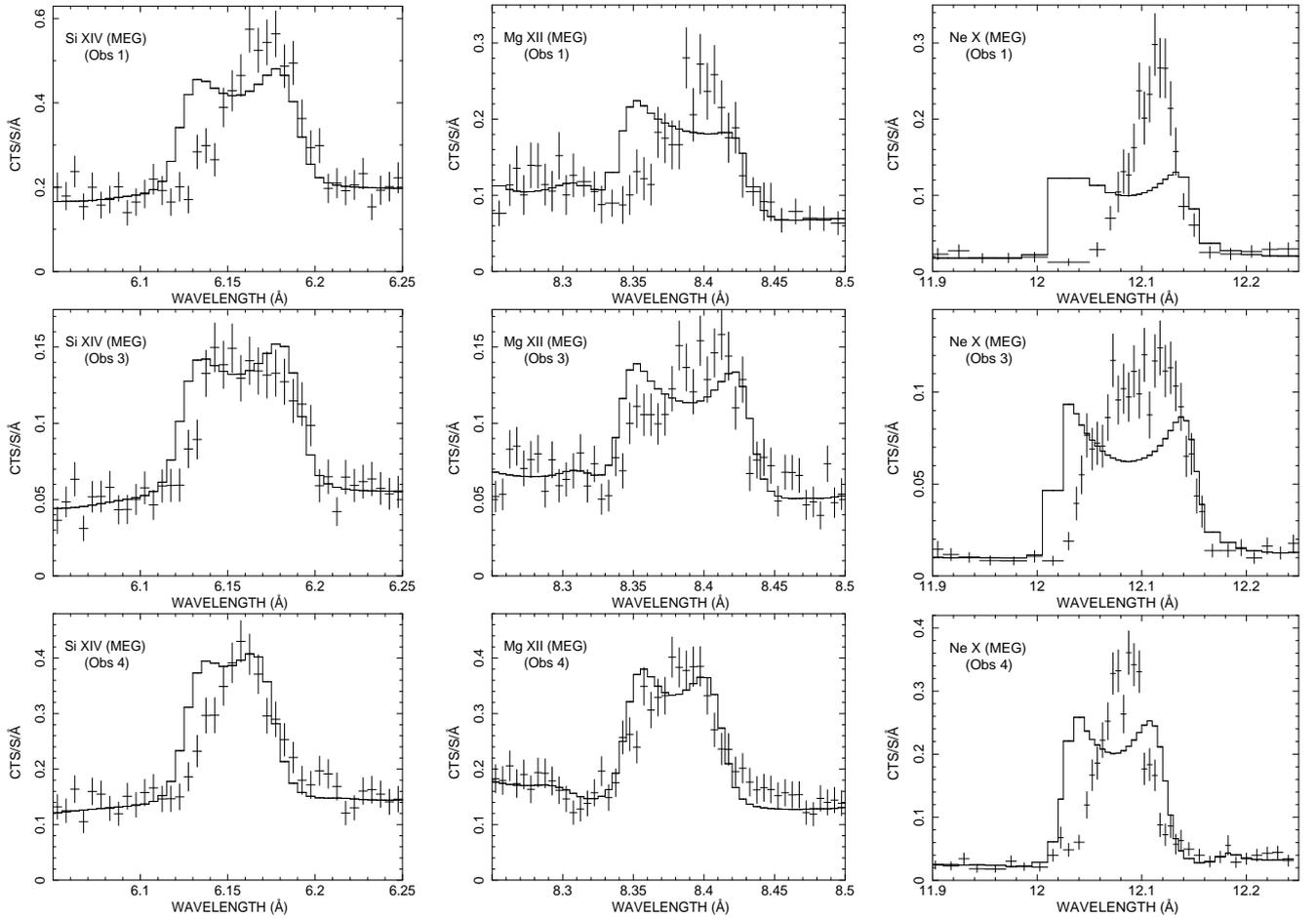

\begin{center}
\includegraphics[width=1.6in,height=2.3in,angle=-90]{fig8a.eps}
\includegraphics[width=1.6in,height=2.3in,angle=-90]{fig8b.eps}
\includegraphics[width=1.6in,height=2.3in,angle=-90]{fig8c.eps}
\includegraphics[width=1.6in,height=2.3in,angle=-90]{fig8d.eps}
\includegraphics[width=1.6in,height=2.3in,angle=-90]{fig8e.eps}
\includegraphics[width=1.6in,height=2.3in,angle=-90]{fig8f.eps}
\includegraphics[width=1.6in,height=2.3in,angle=-90]{fig8g.eps}
\includegraphics[width=1.6in,height=2.3in,angle=-90]{fig8h.eps}
\includegraphics[width=1.6in,height=2.3in,angle=-90]{fig8i.eps}
\end{center}
\caption{The HETG-MEG background-subtracted spectra of \WR and the CSW
model ($\beta = 0.1$; C/He = 0.4) fit near some strong H-like emission
lines (Si XIV 6.18 \AA, Mg XI 8.42 \AA, Ne X 12.13 \AA). The data of
Obs 2 are not shown due to the poor photon statistics that provides no
valuable pieces of information on the gas kinematics (line profiles).
}
\label{fig:lines_meg}
\end{figure*}

\begin{figure*}
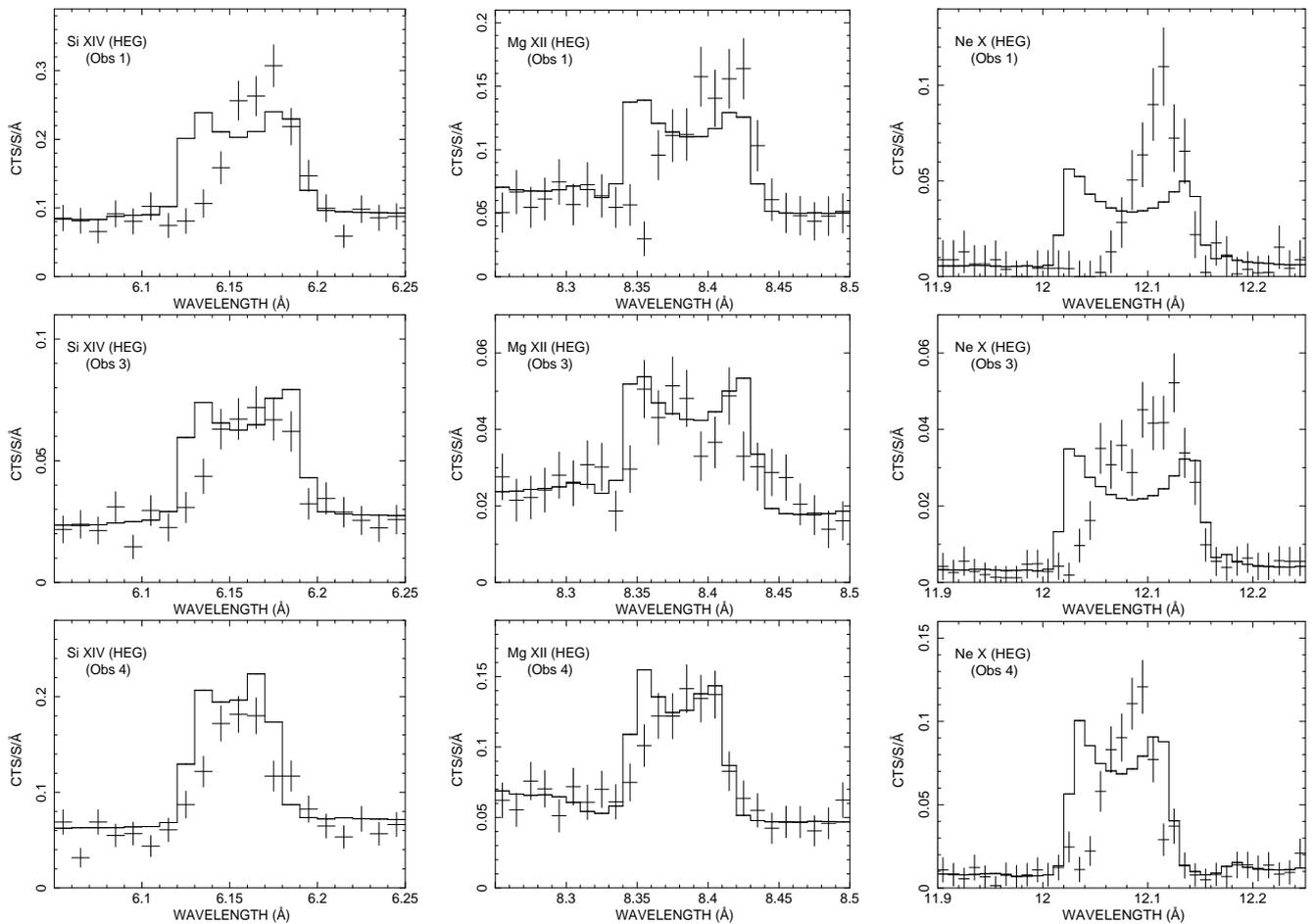

\begin{center}
\includegraphics[width=1.6in,height=2.3in,angle=-90]{fig9a.eps}
\includegraphics[width=1.6in,height=2.3in,angle=-90]{fig9b.eps}
\includegraphics[width=1.6in,height=2.3in,angle=-90]{fig9c.eps}
\includegraphics[width=1.6in,height=2.3in,angle=-90]{fig9d.eps}
\includegraphics[width=1.6in,height=2.3in,angle=-90]{fig9e.eps}
\includegraphics[width=1.6in,height=2.3in,angle=-90]{fig9f.eps}
\includegraphics[width=1.6in,height=2.3in,angle=-90]{fig9g.eps}
\includegraphics[width=1.6in,height=2.3in,angle=-90]{fig9h.eps}
\includegraphics[width=1.6in,height=2.3in,angle=-90]{fig9i.eps}
\end{center}
\caption{The same as in Fig.~\ref{fig:lines_meg} but for the HETG-HEG
background-subtracted spectra of \WR and the same CSW model fit.
}
\label{fig:lines_heg}
\end{figure*}

\begin{figure*}
\begin{center}
\includegraphics[width=\columnwidth]{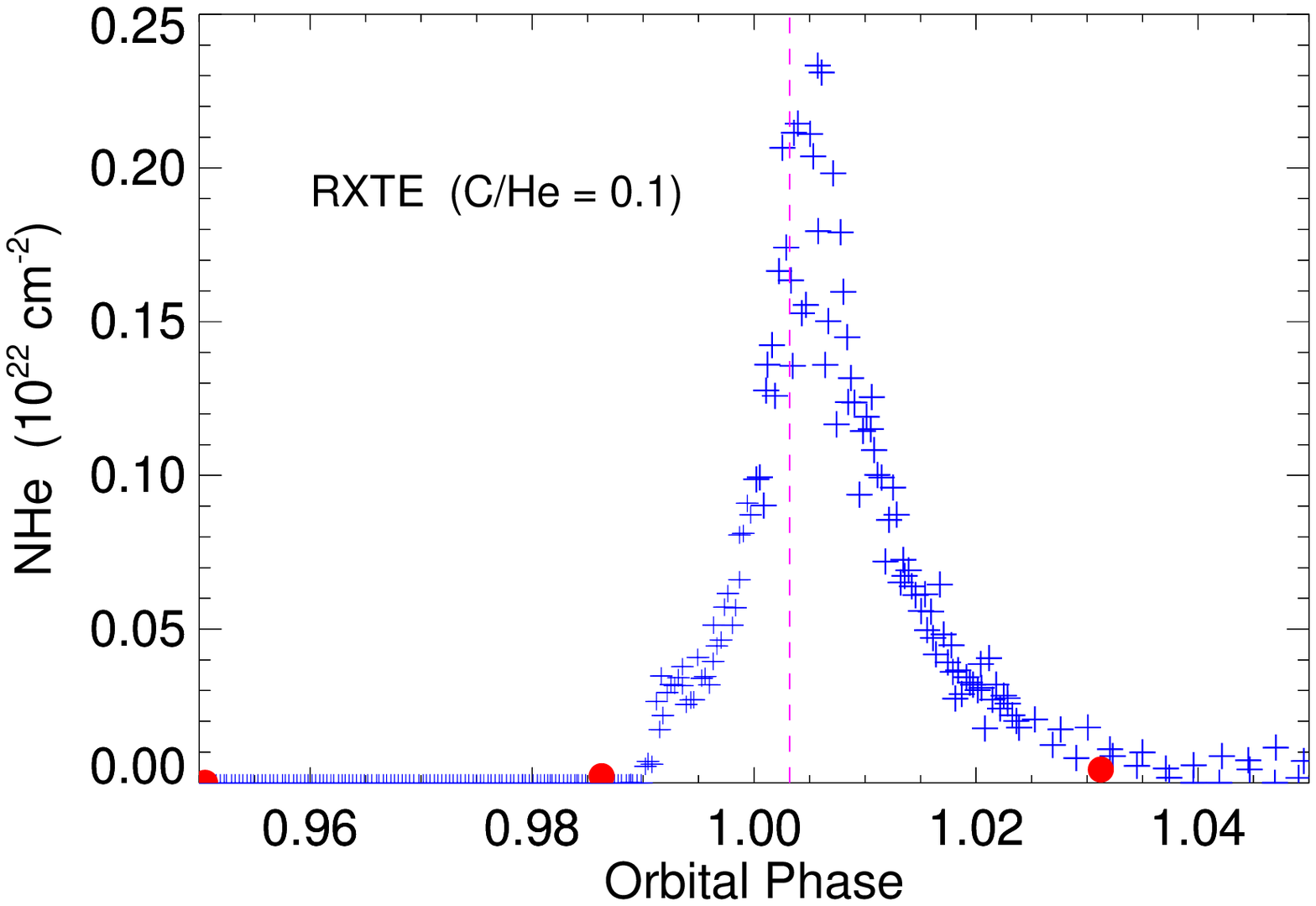}
\includegraphics[width=\columnwidth]{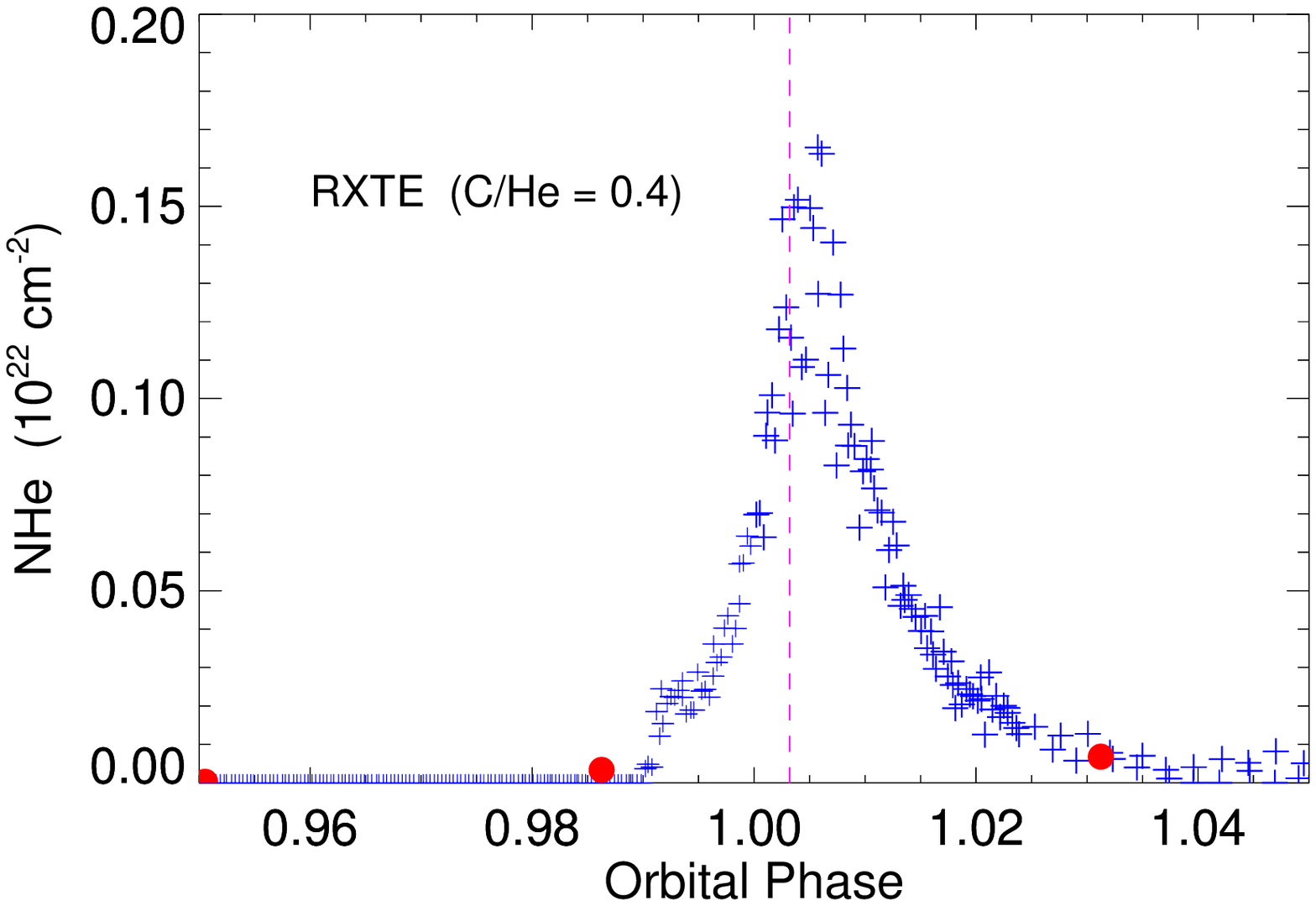}
\includegraphics[width=\columnwidth]{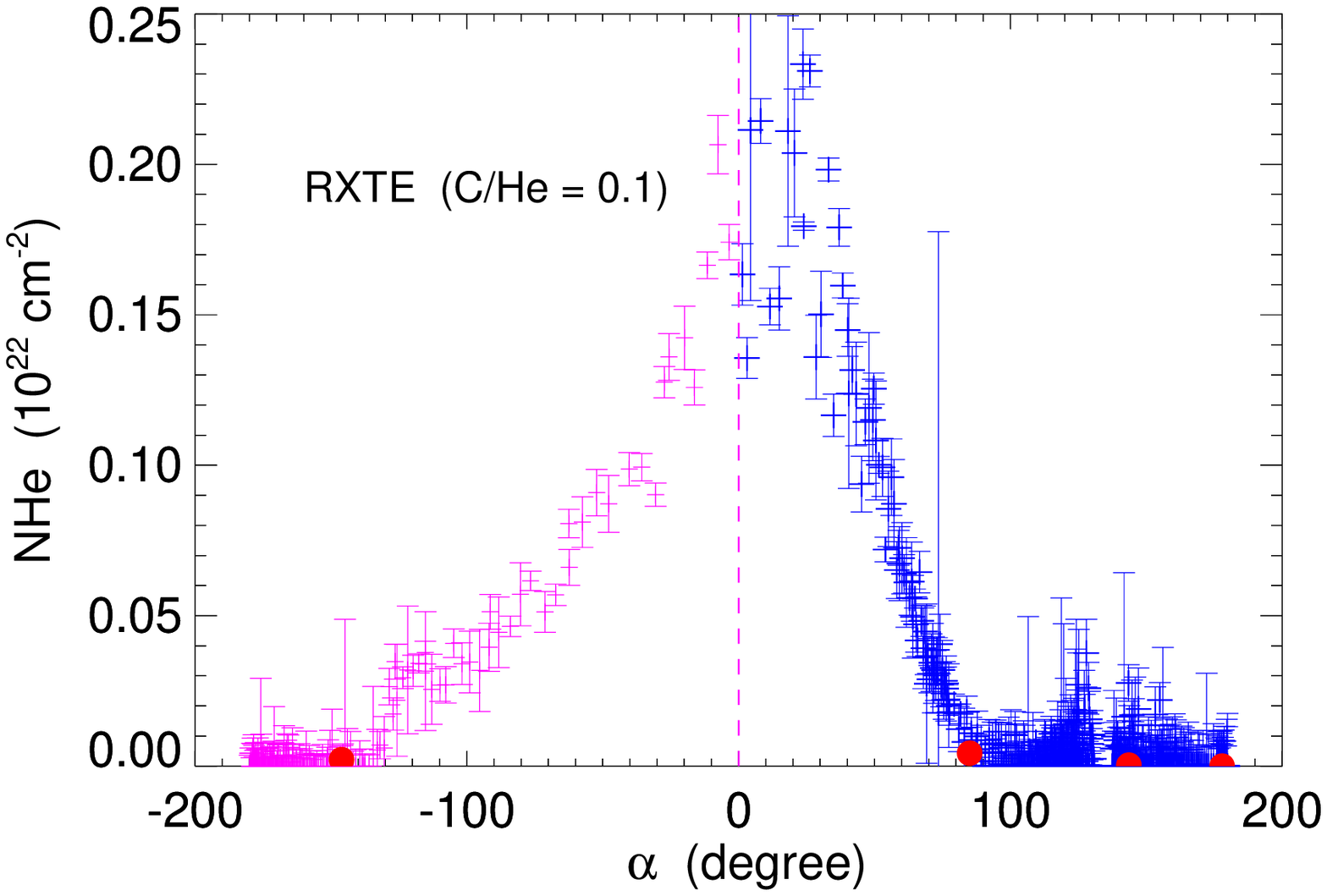}
\includegraphics[width=\columnwidth]{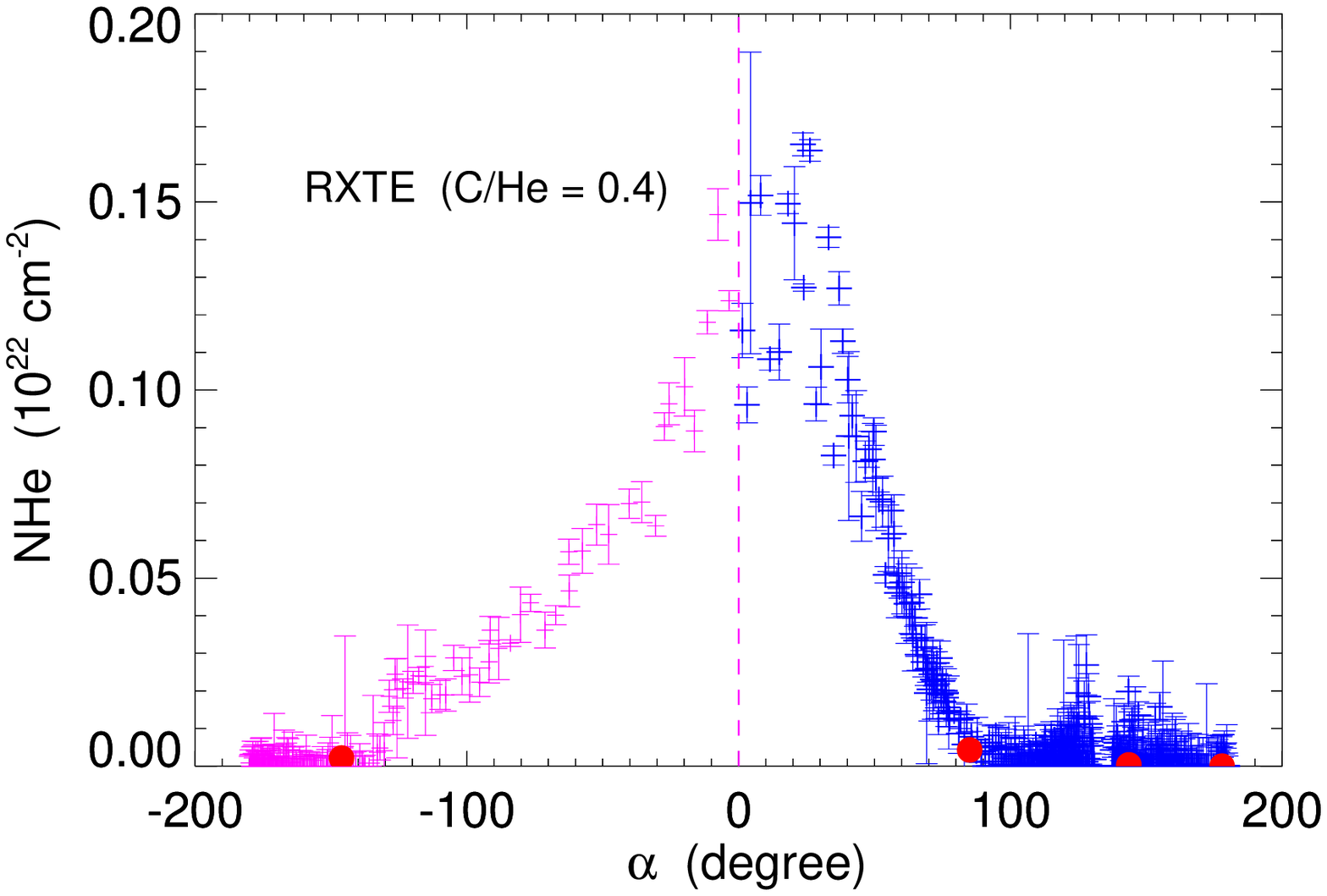}
\includegraphics[width=\columnwidth]{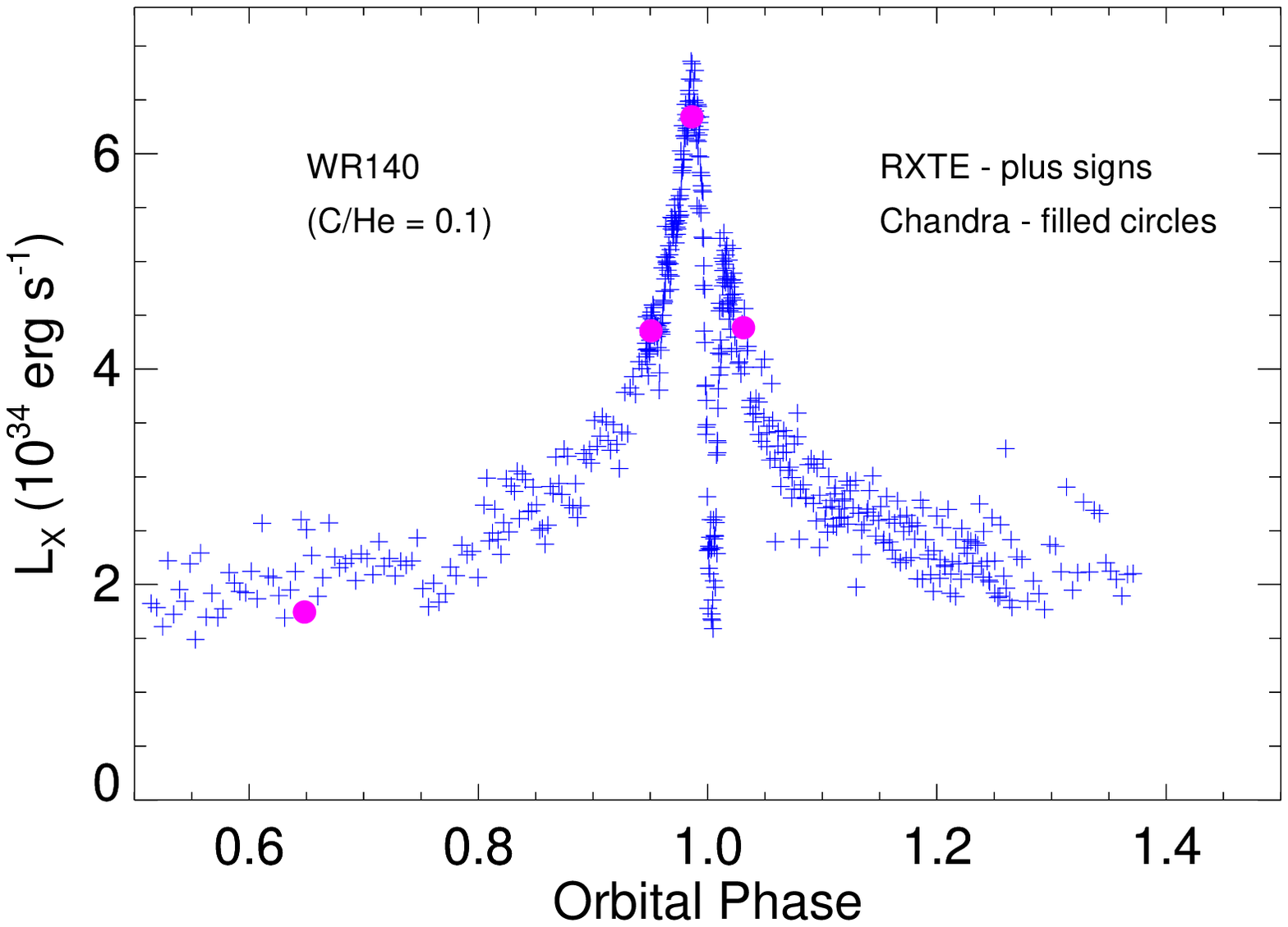}
\includegraphics[width=\columnwidth]{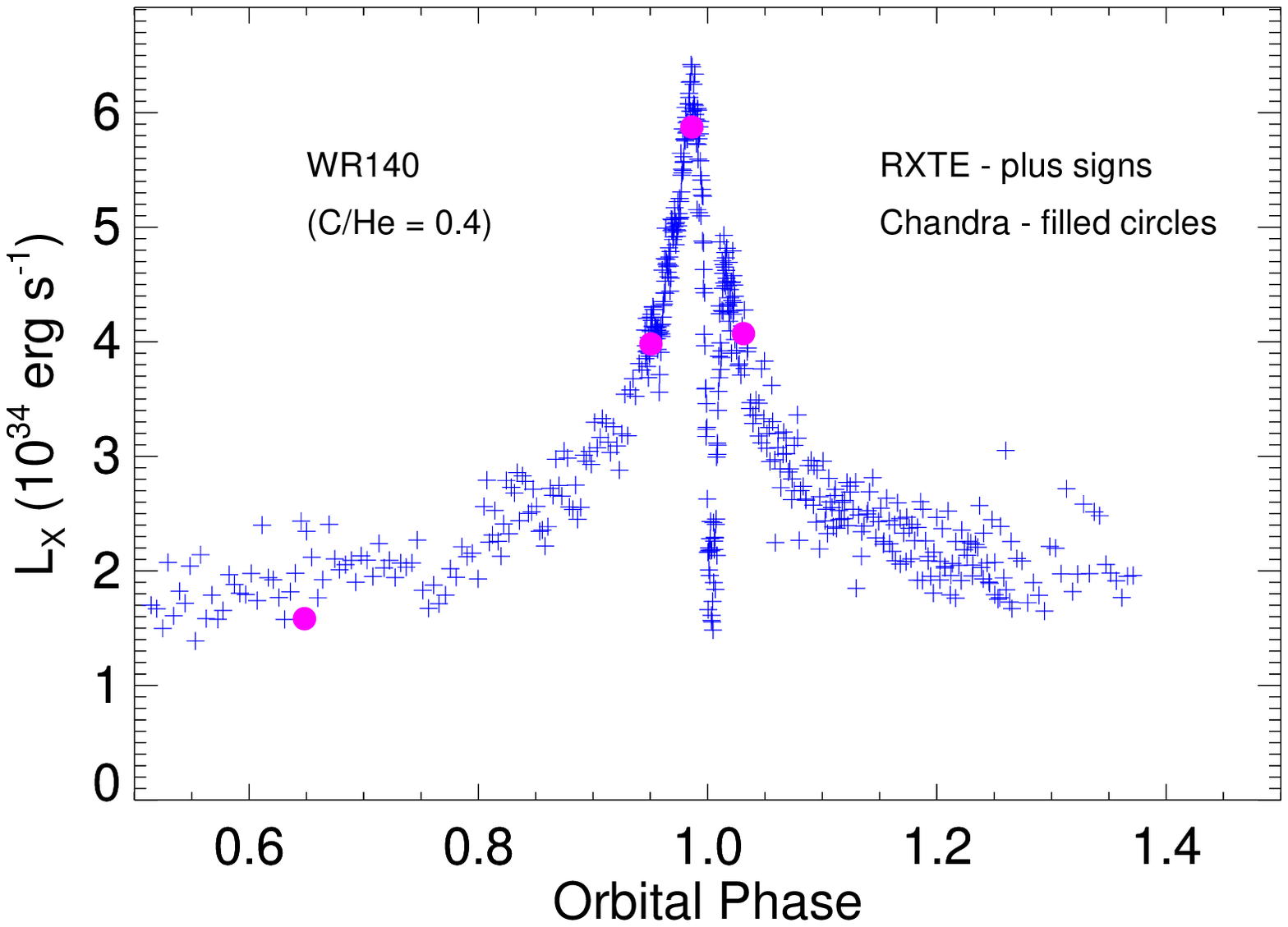}
\end{center}
\caption{
The values of the column density of the `extra' X-ray absorption
(i.e., in excess to that of the stellar winds)
and the X-ray luminosity (0.5 - 10 keV)
as derived from the fit of the CSW model to the \rxte (blue plus
signs)
and \Chandra (red filled circles) spectra of \WR for the
case of different electron and ion temperature ($\beta = 0.1$) and
different chemical composition (C/He = 0.1; 0.4).
The top panels show the X-ray absorption at orbital phases near
periastron (no error bars plotted for clarity).
The middle panels show the X-ray absorption as a function of azimuthal
angle $\alpha$ (see Fig.~\ref{fig:cratun}). Negative values of
$\alpha$ simply denote orbital phases between O-star in front and
WR-star in front (see Fig.~\ref{fig:cratun}).
The vertical dashed line denotes the orbital phase of WR-star in
front.
}
\label{fig:nh_lum}
\end{figure*}

3) Since the gas kinematics of the X-ray emitting plasma is `encoded'
in the spectral line profiles, it is therefore important to check
whether the CSW model could explain/match the observed line profiles
of various ionic species. Similarly to the CSW analysis of the \rxte
spectra (i.e., correspondingly fixed abundances to those from
Table~\ref{tab:fits}), we fitted the high-resolution spectra with 
spectral model that explicitly calculates the line broadening as 
derived in the CSW hydrodynamic simulations. As given in
Section~\ref{sec:csw}, this model also takes into account the stellar 
wind absorption along the line of sight for each parcel of gas in the 
CSW region.
It also allows for an additional absorption component if the stellar
wind absorption is not enough to match the shape of the observed
spectrum. This `extra' absorption component has the same abundances 
as the WR wind.
In \xspec terms  the fitted model reads:
$Spec = wabs(ISM) * wabs(extra) * csw\_lines\_wind$, where
$csw\_lines\_wind$ is our CSW model with line-broadening and stellar
wind absorption along the line of sight taken into account.

For this in-detail check of the CSW picture, we fitted the \Chandra
HEG and MEG spectra adopting one of the `best-fit' CSW models that
successfully represent the \rxte spectra of \WRE. Namely, we used the
CSW model with different electron and ion temperatures for the case of
$\beta = 0.1$.
Some fit results are shown in Figs.~\ref{fig:lines_meg} and
\ref{fig:lines_heg} which illustrate that the CSW
model provides line profiles that in general do {\it not} match well 
the observed line profiles of the strong line features in the X-ray
spectrum of \WRE.
We recall that our CSW \xspec model does not take into account the
thermal line broadening that may in general have some observational
effect, especially, in the case of two-temperature plasma. However, we
see that the theoretical line profiles are in general {\it broader} 
than those observed. Thus, including additional line broadening would 
not improve the correspondence between theory and observations.

Another interesting result is that some small (but not zero) `extra' 
X-ray absorption (i.e., in excess to that of the stellar wind) is
needed to match the shape of the observed spectra at least of 
Obs 1 and Obs 2 (orbital phases 0.9863 and 0.0312, respectively).
To explore further this result, we used the same CSW model (but
switching off the spectral-line broadening) to fit the \rxte spectra of
\WRE. Some fit results are shown in Fig.~\ref{fig:nh_lum}. As seen, 
the fits to the \rxte spectra establish that a considerable X-ray 
absorption in \WR is present at orbital phases near periastron. Most 
importantly, this absorption is in addition to that from the stellar 
winds.

It is worth mentioning that signs of X-ray absorption in `excess' to
that due to the stellar winds were found also from modelling of 
undispersed {\it ASCA} spectra of \WR in the framework of the standard 
CSW picture (see section 4.4.2 in \citealt{zhsk_00}).

\section{Discussion}
\label{sec:discussion}
We carried out a direct modelling of observed X-ray spectra 
(\rxteE, \ChandraE) of \WR in the framework of the {\it standard} 
colliding stellar wind picture in massive WR+O binaries. 
The good correspondence between the results from the \rxte and 
\Chandra spectral fits (Figs.~\ref{fig:rxte_cha_flux},
\ref{fig:norm_fit} and \ref{fig:nh_lum}) 
gives us confidence in the derived CSW 
model results as two of the most important ones are the following. 
First, a considerable reduction of the emission measure of the X-ray
plasma is required at 
orbital phases near periastron in this binary with high orbital 
eccentricity to explain its observed X-ray emission. Second, X-ray 
absorption in excess to that from the stellar winds is present in the 
massive binary \WRE.
On the other hand, a third very important result from modelling the
high-resolution spectra (\ChandraE) of \WRE, adopting the {\it
standard} CSW model, is that such a model does not provide a good
match to the observed profiles of its strong emission lines 
(Figs.~\ref{fig:lines_meg} and \ref{fig:lines_heg}).
We will next discuss their possible implication for the CSW picture 
in massive binaries.

As shown in Section~\ref{sec:fit_results}, the change of the emission
measure can be represented successfully by two terms: one is symmetric
with respect to orbital phase $= 0$ (periastron) and the second term 
is symmetric with respect to the orbital phase the WR stars is in 
front (phase $= 0.00319$): terms $\dot{M} (a)$ and $Scl (\alpha)$,
respectively (see eq,~\ref{eqn:norm}, Figs.~\ref{fig:norm}
and \ref{fig:norm_fit}).
We recall that the $1/a$-dependence of the emission measure on the
binary separation in a massive binary with eccentric orbit is already 
taken into account in the hydrodynamic simulations. Therefore, the
variation of the emission measure in fact indicates a change of the
mass-loss rates. It is generally accepted that the stellar winds of 
massive stars are not uniform, that is they consist of dense clumps
with volume filling factor (clumping) $f < 1$. We could then propose
the following interpretation for the change of the emission measure
symmetric with respect to periastron (i.e., term $\dot{M} (a)$).

Namely, the change of the $\dot{M} (a)$-term over the binary orbit 
(see Fig.~\ref{fig:norm_fit}) could mean that clumps are dissolved in
the CSW region at large binary separations (i.e., near apastron) but 
are not so at small separations (i.e., near periastron) and can easily 
cross the interaction region, thus, not contributing to the X-ray 
emission. The latter is equivalent to smaller {\it effective} 
mass-loss rates as deduced from the spectral fits.

Based on numerical hydrodynamic simulations of `clumpy' CSWs in wide 
binaries,  \citet{pittard_07} concluded that clumps are quickly heated 
up and dissolved in it after crossing the shock fronts of the CSW 
zone. In that study, adiabatic CSW zone was considered in a binary
system with parameters reminiscent of those in \WR at apastron.
It is worth mentioning that dissipative processes may also play a role
for the physics of CSWs, especially, in wide binaries. As shown by the
numerical simulations of \citet{mzh_98}, thermal conduction is more
efficient at large binary separation all other physical parameters 
being the same. Therefore, it will facilitate the clumps to heat up
and evaporate, i.e. dissolve in the interaction region.
Also, we note that the crossing time of the interaction region for a 
clump is longer at large binary separations (i.e., near apastron) 
since the width of the interaction region is proportional to the 
binary separation. As a result, the clump will experience the 
`evaporation force' for longer time which increases its chances to 
being dissolved in the interaction region.

As to the $Scl (\alpha)$-term in the variable emission measure (that 
symmetric with respect to orbital phase the WR star is in front), we 
think it could be considered qualitatively in conjunction with results 
on the `extra' X-ray absorption in \WRE.  We note that the latter has 
its maximum values near the same orbital phase (WR in front) but it is 
asymmetric with respect to it (see Fig.~\ref{fig:nh_lum}). It may seem 
speculative, but could it be that all this (the necessity of the $Scl
(\alpha)$-term and the presence ot `extra' X-ray absorption) is a hint
on {\it non-spherically-symmetric} 
stellar wind(s) in the massive binary \WRE ?

And, could it be that a similar sign might also be coming from the 
fact that the {\it standard} CSW model (i.e. considering interaction 
of spherically-symmetric stellar winds) failed to produce a good match 
to the observed line profiles in \WR (see Figs.~\ref{fig:lines_meg} 
and \ref{fig:lines_heg})?

In fact, we ran models with different values of orbital inclination
(e.g., $i = 50, 70, 80$ deg; see Section~\ref{sec:star}) but they did
not provide a better match to the observed line profiles. The same is
valid for our modelling with increased value of the ram-pressure ratio
$(\Lambda = 39$), which in general may produce line profiles with
smaller widths.
We also tried to mimic a non-spherical absorber (e.g., similar to some 
disk-like structure as that proposed from radio observations of \WR;
see section 4 in \citealt{whi_bec_95}) by taking into account {\it
only} the X-ray emission `below' or `above' the orbital plane, or from
a certain sector of the CSW cone. However, this toy-play modelling
did not improve the correspondence between the CSW model line profiles
and those observed.

And, we note that the \Chandra observations Obs 1 and Obs 3 have been
carried out at orbital phases that correspond to very similar values
of the azimuthal angle: $\alpha = 146$ (Obs 1) and $\alpha = -144$ 
(Obs 3; note that the case with a given value of $\alpha$ is
equivalent to that with the value $\alpha = 360 - \alpha$ and is 
equivalent to that with the value $-\alpha$; see Fig.~\ref{fig:cratun}). 
This suggests that the 
resultant line profiles should be almost identical in the case of CSW 
region produced in collision of two spherically-symmetric stellar 
winds. However, the observed line profiles (shapes, widths) are quite 
different between Obs 1 and Obs 3 
(e.g., see Figs.~\ref{fig:lines_meg} and \ref{fig:lines_heg}).

Thus, it is our understanding that a next step in development of more
realistic CSW models could be to consider CSW models with 
non-spherically-symmetric (e.g., axisymmetric) 
stellar winds. We have to keep in mind that such a model is definitely 
a three-dimensional (3D) model. Exploring its parameter space is a 
heavy task since even if
only one of the winds is axisymmetric the resultant interaction region 
will depend on the orientation in space of the axis of symmetry of the 
stellar wind that will be changing over the orbit. Nevertheless, it is 
worth to try, we think.

It is important to mention another feature that adds to
complexity of modelling CSWs in massive binaries. As discussed in
\citet{zh_etal_20} (see section 4.3 therein), if the winds of massive
stars are {\it just} `clumpy' (with volume filling factor $f < 1$) it 
is not clear how clumps manage to collide and form the interaction
region in a massive binary.
It is thus reasonable to assume that stellar winds of massive
stars are {\it two-component} flows. The more massive component
in the flow
consists of dense clumps that may occupy even a very small fraction 
of the stellar wind volume (e.g., with $f = 0.1 - 0.25$, see
\citealt{hamann_19}, \citealt{sander_19}, and references therein).
And, there is a low-density component that fills in the rest of the 
volume. In such a physical picture, the low-density wind components 
of both massive stars in the binary interact and set up the `seed' 
CSW region. The more massive components of the stellar winds (the 
clumps) then interact with it and provide the strong X-ray emission 
observed from wide CSW binaries.
Note that modelling such a two-component wind interaction will also
explicitly require  3D hydrodynamic simulations.

Finally, we recall that the physical picture of two-component stellar
winds in massive stars seems to get some support also from the 
relatively low level of X-ray emission from close WR$+$O binaries that 
may likely arise from {\it adiabatic} CSW shocks of the low-density 
components of the stellar winds \citep{zh_12}.

\section{Conclusions}
\label{sec:conclusions}
The basic results and conclusions from our modelling of the \Chandra 
and \rxte X-ray spectra of the massive binary \WR {\it in the 
framework of the standard colliding stellar wind picture} are 
as follows.

(i) 
CSW model spectra match well the shape of the observed X-ray
spectrum of \WRE.  Models with partial electron heating at the
shock fronts (different electron and ion temperatures) with
$\beta = T_e / T < 0.4$ ($T_e$ is the electron temperature and 
$T$ is the mean plasma temperature) are a better representation 
of the X-ray data than those with complete temperature equalization.

(ii) 
A considerable decrease is found for the emission measure of the X-ray
plasma in the CSW region at orbital phases near periastron. This is
equivalent to variable effective mass-loss rates over the binary
orbit.

(iii) 
A considerable X-ray absorption in \WR is present at orbital phases 
near periastron, as this absorption is in {\it excess} to that from 
the stellar winds.

(iv) 
Fits to the high-resolution spectra showed that the {\it standard} 
CSW model provides line profiles that in general do {\it not} match 
well the observed line profiles of the strong line features in the 
X-ray spectrum of \WRE.

(v) 
To explain these findings, we propose a qualitative picture for wide
massive binaries whose orbits have high eccentricity as is the case of
\WRE. In this picture, stellar winds are `clumpy'. Due to some
dissipative processes (e.g., thermal conduction), clumps are easily 
dissolved in the CSW region at large binary separations (i.e., near 
apastron) but are not so at small separations (i.e., near periastron) 
and can easily cross the interaction region, thus, not contributing 
to the X-ray emission. The latter is equivalent to variable (e.g.,
smaller near periastron) {\it effective} mass-loss rates over the 
binary orbit. On the other hand, we think that {\it future} developing
of  CSW models with {\it non-spherically-asymmetric} stellar winds 
might be the way to resolve the
mismatch between the theoretical and observed X-ray line profiles 
and explain the excess X-ray absorption near periastron in \WRE, the
{\it prototype} of CSW massive binaries.

\section*{Acknowledgements}
The author, Svetozar Zhekov, dedicates this study to his dear friend
and colleague Artyom Myasnikov.
This research has made use of data and/or software provided by the
High Energy Astrophysics Science Archive Research Center (HEASARC),
which is a service of the Astrophysics Science Division at NASA/GSFC
and the High Energy Astrophysics Division of the Smithsonian
Astrophysical Observatory. 
This research has made use of the NASA's Astrophysics Data System, and
the SIMBAD astronomical data base, operated by CDS at Strasbourg,
France.
The author acknowledges financial support from
Bulgarian National Science Fund grant DH 08 12.
The author thanks an anonymous referee for 
valuable comments and suggestions.

\section*{Data Availability}

The X-ray data underlying this research are {\it
public} and can be accessed as follows.
The \Chandra data sets can be downloaded from the \Chandra X-ray
observatory data archive
\url{https://cxc.harvard.edu/cda/} by typing in the target name (\WRE)
in the general search form \url{https://cda.harvard.edu/chaser/}.
The \rxte data sets can be downloaded from the \rxte observatory  part 
of the NASA's High Energy Astrophysics Science Archive
\url{https://heasarc.gsfc.nasa.gov/docs/archive.html} by typing in the
object name (\WRE) in the general search form 
\url{https://heasarc.gsfc.nasa.gov/cgi-bin/W3Browse/w3browse.pl}.
Some details on the data reduction are given in
Section~\ref{sec:data}.



\bibliographystyle{mnras}
\bibliography{wr140} 


\appendix
\section{Some results from the CSW model fits to the \rxte spectra}
\label{app}

In Table~\ref{tab:chi2}, we provide in detail the reduced $\chi^2$
values for the CSW model fits to the \rxte spectra that could be used
in supplementary analysis.

\begin{table*}

\caption{Reduced $\chi^2$ values for the fits to the \rxte spectra}
\label{tab:chi2}

\begin{center}
\begin{tabular}{rcccccc}
\hline
\multicolumn{1}{c}{ } &
\multicolumn{3}{c}{C / He = 0.1}  &
\multicolumn{3}{c}{C / He = 0.4}  \\
\multicolumn{1}{c}{$\beta$} & 
\multicolumn{1}{c}{nominal} &
\multicolumn{1}{c}{$\dot{M} (a)$} & \multicolumn{1}{c}{$Scl (\alpha)$}
&
\multicolumn{1}{c}{nominal} &
\multicolumn{1}{c}{$\dot{M} (a)$} & \multicolumn{1}{c}{$Scl (\alpha)$}
\\
\hline
   0.001 & 1.052566 & 0.895815 & 0.889423 & 1.065246 & 0.893618 &
0.886700 \\
    0.05 & 1.049501 & 0.892688 & 0.886263 & 1.063540 & 0.891672 &
0.884810 \\
     0.1 & 1.049102 & 0.893042 & 0.886711 & 1.064347 & 0.892993 &
0.886250 \\
     0.2 & 1.060226 & 0.905153 & 0.898958 & 1.077858 & 0.906784 &
0.899959 \\
     0.3 & 1.082457 & 0.930499 & 0.924067 & 1.102827 & 0.935067 &
0.928148 \\
     0.4 & 1.114851 & 0.969903 & 0.963671 & 1.141406 & 0.988448 &
0.981739 \\
     0.5 & 1.156347 & 1.024172 & 1.018375 & 1.191568 & 1.045426 &
1.039512 \\
     0.6 & 1.206856 & 1.093312 & 1.088236 & 1.252401 & 1.132138 &
1.122070 \\
     0.7 & 1.266157 & 1.175927 & 1.172277 & 1.322603 & 1.223760 &
1.219941 \\
     0.8 & 1.334866 & 1.272470 & 1.269603 & 1.400209 & 1.332027 &
1.329423 \\
     0.9 & 1.413030 & 1.380978 & 1.379454 & 1.485615 & 1.451425 &
1.449769 \\
     1.0 & 1.498292 & 1.498346 & 1.498398 & 1.577701 & 1.577820 &
1.577928 \\

\hline

\end{tabular}
\end{center}

{\it Note}. The total reduced $\chi^2$ (dof $= 22080$) for the entire
set of 552 \rxte spectra for each value of parameter $\beta = T_e /
T$ ($T_e$ is the electron temperature and $T$ is the mean plasma 
temperature) and both basic sets of abundances considered in this study
(C / He = 0.1; C / He = 0.4, see Section~\ref{sec:fit_results}).
Columns `nominal', $\dot{M} (a)$ and $Scl (\alpha)$ present results
from the CSW model fits for the cases with the nominal
stellar-wind parameters, reduced mass-loss rates
as described by terms $\dot{M} (a)$ and $Scl (\alpha)$, respectively 
(see eq.~\ref{eqn:norm} and the details in
Section~\ref{sec:fit_results}). 
These results are shown in Fig.~\ref{fig:chi2}.

\end{table*}

\bsp	
\label{lastpage}
\end{document}